\documentclass[preprint]{elsarticle}
\usepackage{graphicx}
\usepackage{dcolumn}
\usepackage{bm}
\usepackage{amssymb}
\usepackage{amsmath}

\begin{document}

\def\vri{\vec{r}_{i}} \def\vrj{\vec{r}_{j}} \def\rij{r_{ij}}
\def\vrij{\vec{r}_{ij}} \def\drij{\hat{r}_{ij}}
\def\vdr{\delta\vec{r}} \def\dr{\delta{r}} \def\s{\hat{s}}

\title{Jamming II: Edwards' statistical mechanics of random packings of hard spheres}

\author[rvt]{Ping Wang}
\author[rvt]{Chaoming Song}
\author[rvt]{Yuliang Jin}
\author[rvt]{Hern\'an A. Makse\corref{cor1}}
\ead{hmakse@lev.ccny.cuny.edu}
\cortext[cor1]{Corresponding author}
\address[rvt]{Levich Institute and Physics Department, City
  College of New York, New York, NY 10031, US}


\begin{abstract}
{\bf


The problem of finding the most efficient way to pack spheres has an
illustrious history, dating back to the crystalline arrays conjectured
by Kepler and the random geometries explored by Bernal in the 60's.
This problem finds applications spanning from the mathematician's
pencil, the processing of granular materials, the jamming and
glass transitions, all the way to fruit packing in every grocery.
There are presently numerous experiments showing that the loosest
way to pack spheres gives a density of $\sim55$\% (named random
loose packing, RLP) while filling all the loose voids results in a
maximum density of $\sim 63-64$\% (named random close packing,
RCP). While those values seem robustly true, to this date there is
no well-accepted physical explanation or theoretical
prediction for them.
Here we develop a common framework for understanding the random
packings of monodisperse hard spheres
whose limits can be
interpreted as the experimentally observed RLP and RCP. The reason
for these limits arises from a statistical picture of jammed
states in which the RCP can be interpreted as the ground state of
the ensemble of jammed matter with zero compactivity, while the
RLP arises in the infinite compactivity limit.
We combine an extended statistical mechanics approach 'a la
Edwards' (where the role traditionally played by the energy and
temperature in thermal systems is substituted by the volume and
compactivity) with a constraint on mechanical stability imposed by
the isostatic condition.  We show how such approaches can bring
results that can be compared to experiments and allow for an
exploitation of the statistical mechanics framework.  The key
result is the use of a relation between the local Voronoi volumes
of the constituent grains (denoted the volume function) and the
number of neighbors in contact that permits to simple combine
the two approaches to develop a theory of volume fluctuations in
jammed matter.  Ultimately, our results lead to a phase diagram
that provides a unifying view of the disordered hard sphere
packing problem and further shedding light on a diverse spectrum
of data, including the RLP state.  Theoretical results are well
reproduced by numerical simulations that confirm the essential
role played by friction in determining both the RLP and RCP
limits.
The RLP values depend
on friction, explaining why varied experimental results can be
obtained.



}

\end{abstract}
\maketitle

\section{Introduction}


Filling containers with balls is one of the oldest mathematical
puzzles known to scientists. The study of disordered sphere packings raises an interesting problem: How efficient and uniform will spheres pack if assembled randomly? This problem has an important application in the jamming phenomenon, which takes place in particulate systems
where all particles are in close contact with one another. The study of jammed granular media offers
unexpected challenges in physics since the equilibrium statistical
mechanics fail for these out-of-equilibrium systems. The
goal of the present study is to develop an ensemble theory of
volume fluctuations to describe the statistical mechanics of
jammed matter with the aim of shedding light to the long-standing
problem of characterizing random sphere packings.

\subsection{Sphere packing problem}

 The study of sphere packing problem started
four centuries ago when Johannes Kepler conjectured that the most
efficient arrangement of spheres is the FCC lattice (an important
part of the 18th problem proposed by Hilbert in 1900). Even though
this is a tool used for centuries in fruit markets around the
globe, nearly 400 years passed before this conjecture was
considered a mathematical proof, which has been developed only
recently by Hales \cite{hales} in a series of articles covering
250 pages supplemented by 3Gb of computer code to determine the
best ordered packing through linear programming. The difficulty
arises since in 3d it is not enough to look at the packing of one
cell, but it is necessary to consider several Voronoi cells at
once. That is, the packing that minimize the volume locally (the
dodecahedron) does not tile the system globally.  Such a situation
does not arise in 2d, where the hexagonal packing minimizes the
volume locally and globally; the equivalent of Kepler
conjecture in 2d, which is also known as Thue's theorem, was
proved long ago \cite{thue92,thue10}.

The analogous problem for disordered packings has also an
illustrious but unfinished history. This problem was initiated by
the pioneering work of Bernal in the 1960's \cite{bernal}
(although earlier attempts can be found). The traditional view
states that \cite{anonymous} ``packings of spherical particles
have been shaken, settled in different fluids and kneaded inside
rubber balloons and all with no better results than a maximum
density of 63\%''.  This is the so-called random close packing RCP
limit \cite{bernal,anonymous,scott,finney,berryman}.  On the other
hand, other experiments have shown that densities as low as 55\%
can be obtained in random loose packings, RLP
\cite{scott,berryman,onoda}. 
While the two limits seem reproducible in various experiments and computer simulations, a mathematically-rigorous definition is still unavailable. Indeed it was conjectured
\cite{torquato} that the
RCP conception is mathematically ill-defined and should be replaced by the maximally random jammed
(MRJ) conception in terms of an ensemble of order parameters.

To this date there is no well-accepted
physical explanation of this phenomenon, no well-accepted
theoretical prediction of such density values and heated debates
are still found in the literature regarding the existence of
rigorous definitions of the RCP and RLP, the uniqueness of the RCP
state and the nature of their state of randomness.

\subsection{Understanding Jamming}
One of the most important physical applications of the sphere
packing problem is to understand the intrinsic geometrical
structures of jammed systems. A great deal of
research has been devoted by physicists to the properties of
packings of granular materials and other jammed systems such as
compressed emulsions, dense colloids and glasses
\cite{unifying,powders}, due to their ``out-of-equilibrium"
nature. Generally speaking, there are three interrelated
approaches to understand the nature of the jammed state of matter
and its ensuing jamming transition :

{\bf (a) Statistical mechanics of jammed matter}: The jammed phase
is described with the ensemble of volume and force fluctuations
proposed by Edwards
\cite{edward,edwards2,angoricity,blumenfeld,dauchot,ciamarra,chicago2,sdr,swinney}.
Ideas from the physics of glasses provide interesting
cross-fertilization between the jamming transition and the glass
transition \cite{ck,barrat,mk,coniglio}.

{\bf (b) Critical phenomena of deformable particle}: jamming is
seen as a critical point at which observables such as pressure,
coordination number and elastic moduli behave as power-law near
the critical volume fraction, $\phi_c$,
\cite{liu1,makse,ohern,ohernprime,zhang,silbert,weitz,kertesz,vanhecke1}.
The jammed state is modelled as granular media composed of soft
particles, typically Hertz-Mindlin \cite{zhang,landau,mindlin}
grains under external pressure and emulsions compressed under
osmotic pressure \cite{brujic}, as they approach the jamming
transition from the solid phase above the critical density: $\phi
\to \phi_c^+$.

{\bf (c) Hard-sphere glasses}: In parallel to studies in the field of
jamming there exists a community attempting to understand the packing
problem approaching jamming from the liquid phase
\cite{parisi,lubachevsky,skoge}, that is, $\phi \to \phi_c^{-}$.
Here, amorphous jammed packings are seen as infinite pressure glassy
states \cite{parisi,kurchan}. Therefore, the properties of the jamming
transition are intimately related to those of the glass transition
\cite{parisi}.

A great deal of effort has been devoted to the study of jamming
applying these approaches.  A large body of experiments and
simulations have fully characterized the jammed state in the critical
phenomena framework of soft-particles. The hard-sphere field studies
frictionless amorphous packings interacting with hard-core normal
forces only.  Their results are a particular case of the more general
problem of jammed matter composed of frictional and frictionless hard
spheres as treated in the present work.

Statistical mechanics studies of jammed matter were initiated in
1989 with the work of Sir Sam Edwards. It was first recognized
that the main theoretical difficulty to develop a statistical
formulation results from the lack of well-defined conservation
laws on which an ensemble description of jammed matter could be
based. Unlike equilibrium statistical mechanics where energy
conservation holds, granular matter dissipates energy through
frictional inter-particle forces.  Therefore, it is doubtful that
energy in granular matter could describe the microstates of the
system and a new ensemble needs to be considered.

Stemming from the fact that it is possible to explore different
jammed configuration at a given volume in systematic experiments
\cite{chicago2,bideau,sdr,swinney}, Edwards proposed the
volume ensemble (V-ensemble) as a replacement of the energy
ensemble in equilibrium systems.  A simple experiment is merely
pouring grains in a fixed volume and applying perturbations, sound
waves or gentle tapping, to explore the configurational changes.
Concomitant with a given network geometry there is a distribution
of contact forces or stresses in the particulate medium. This
means that the V-ensemble must be supplemented by the force or
stress ensemble (F-ensemble) determined by the contact forces
\cite{angoricity,bulbul1,blum1} for a full characterization of
jammed matter.

Following this theoretical framework, a large body of experiments,
theory and simulations \cite{edward,edwards2,angoricity,blumenfeld,dauchot,ciamarra,chicago2,sdr,swinney,aste,cruz,E8,brujic} have focused on the study of volume
fluctuations in granular media.
Statistical studies have been concerned with testing for the
existence of thermodynamics quantities such as effective
temperatures and compactivity as well as challenging the
foundations based on the ergodic hypothesis or equal probability
of the jammed states. It is recognized that thermodynamic analogies may illuminate
methods for attempting to solve certain granular problems, but
will inevitably fail at some point in their application. The mode
of this failure is an interesting phenomenon, as the range of
phenomena for which ergodicity of some kind or other will apply or
not is an interesting question.  This has been illustrated by the
compaction experiments of the groups of Chicago, Texas, Rennes and
Schlumberger \cite{chicago2,bideau,sdr,swinney}. They
have shown that reversible states exist along a branch of
compaction curve where statistical mechanics is more likely to
work. Conversely, experiments also showed a branch of
irreversibility where the statistical framework is not expected to
work.
Poorly consolidated formations, such as a sandpile, are
irreversible and a new ``out-of-equilibrium'' theory is required
to describe them.

In this paper we will describe the reversible states which are
amenable to the ergodic hypothesis while the unconsolidated
irreversible states will be addressed in a future work.
Unfortunately, there is no first principle derivation of the granular
statistical mechanics analogous to the Liouville theorem for
equilibrium systems \cite{landau-stat-mech}. Therefore, advancing the
statistical mechanics of granular matter requires well-defined
theoretical predictions that can be tested experimentally or
numerically.
While the possibility of a thermodynamic principle describing
jammed matter is recognized as a sensible line of research, the
problem with the statistical approach is that after almost 20
years there are no practical applications yet. We require
predictions of practical importance such as equations of state
relating the observables: pressure, volume, coordination number,
entropy, etc, that may lead to new phenomena to be discovered and
tested experimentally that will allow for a concrete exploitation
of the thermodynamic framework.


\subsection{Objectives}

Here, we explore this problem by developing a systematic study of
the V-ensemble of jammed matter. Our systems of interest are
primarily packings of granular materials, frictional colloids,
infinitely rough grains and frictionless droplets mimicking
concentrated emulsions.
We attempt to answer the following questions using statistical
mechanics methods:

--- What is a jammed state?

--- What are the proper state variables to describe the granular
 system at the ensemble level?



--- What are the ensuing equations of state to relate the different
 observables?  For instance, the simplest equilibrium thermal system,
 the ideal gas, is described by pressure, volume and temperature
 $(p,V,T)$ through $pV = nRT$. Is it possible to
 identify the state variables for jammed matter which can describe the
 system in a specified region of the phase space?

--- Which are the states with the maximum and minimum entropy (most and
least disordered)?


--- Can we define a "compactimeter" to measure compactivity?


--- Can we provide a statistical interpretation of RLP and RCP, and
use the ensemble formalism to predict their density values?

To answer these questions, we illuminate a diverse spectrum of data on sphere packings through a
statistical theory of disordered jammed systems. Our results ultimately lead
to a phase diagram providing a unifying view of the disordered sphere
packing problem. An extensive program of computer simulations of frictional and
frictionless granular media tests the predictions and assumptions of
the theory, finding good general agreement.  The phase diagram
introduced here allows the understanding of how random packings fill
space in 3d for any friction, and eventually can be extended as a
mathematical model for packings in 2d and in higher dimensions,
anisotropic particles and other systems.  While physical application
may be limited to 2d or 3d, higher dimensions find application in the
science of information theory, where data transmission and encryption
find use for packing optimization.


\subsection{Outline}

This paper is the second installment of a series of papers dealing
with different aspects of jammed matter from a statistics mechanics
point of view.  In a companion paper, Jamming I \cite{jamming1}, we
describe the definition of the volume function based on the Voronoi
volume of a particle and its relation to the coordination number.

In the present paper, Jamming II, we describe calculations leading to
the phase diagram of jammed matter and numerical simulations to test
the theoretical results; a short version of the present paper has been
recently published in \cite{jamming2}. Jamming III \cite{jamming3}
describes the entropy calculation to characterize randomness in jammed
matter and a discussion of the V-F ensemble.  Jamming IV
\cite{jamming4} studies the distribution of volumes from the
mesoscopic ensemble point of view and a generalized z-ensemble to
understand the distribution of coordination number.  Jamming V
\cite{jamming5} details the calculations for jamming in two dimensions
at the mesoscopic level.
The theory could be also extended to bidisperse
\cite{Max} and high-dimensional systems \cite{van-Meel}.



The outline of the present paper is as follows: Section
\ref{statistical-mechanics} explain the basis of the Edwards
statistical mechanics.  Section \ref{volume-function} summarizes
the results of Jamming I regarding the calculation of the volume
function. Section \ref{definition} describes the isostatic
condition that defines the ensemble of jammed matter through the
$\Theta_{\rm jam}$ function, and explains the difference between
the geometrical, $z$, and mechanical, $Z$, coordination number,
which is important to define the canonical partition function.
Section \ref{dos} defines the density of states in the partition
function. Section \ref{phase-diagram} explains the main results of
this paper in the equations of state (\ref{state1}) and
(\ref{state2}), and the phase diagram of Fig. \ref{phasea}.
Finally, Section \ref{simulations} explains the numerical studies
to test the theoretical predictions.

\section{Statistical mechanics of jammed matter}
\label{statistical-mechanics}

Conventional statistical mechanics uses the ergodic hypothesis to
derive the microcanonical and canonical ensembles, based on the
quantities conserved, typically the energy $E$
\cite{landau-stat-mech}.  Thus the entropy in the microcanonical
ensemble is $S(E) = k_B \log \int \delta (E-{\cal H}(p,q)) dp dq$,
where ${\cal H}(p,q)$ is the Hamiltonian.  This becomes the canonical
ensemble with $\exp [- {\cal H} (\partial S/\partial E)]$.


The analogous development of a statistical mechanics of granular and
other jammed materials presents many difficulties.  Firstly, the
macroscopic size of the constitutive particles forbids the equilibrium
thermalization of the system. Secondly, the fact that energy is
constantly dissipated via frictional interparticle forces further
renders the problem outside the realm of equilibrium statistical
mechanics due to the lack of energy conservation.  In the absence of
energy conservation laws, a new ensemble is needed in order to
describe the system properties.

Following this theoretical perspective, Edwards proposed the
statistical mechanics of jammed matter and interest in the problem of
volume fluctuations has flourished \cite{edwards2}.
The central concept is that of a volume function ${\cal W}$
replacing the role of the Hamiltonian in describing the
microstates of the system in the V-ensemble and the stress
boundary $$\Pi_{ij} = \int{\sigma_{ij}dV}$$ with the stress
$\sigma_{ij}= 1/(2V) \sum_c f_i^c r_j^c$ describing the F-ensemble
 \cite{angoricity,bulbul1,blum1}, where $f_i^c, r_i^c$ are the force and position at contact $c$.
For simplicity only the isotropic case is described. Thus, only
the pressure $\sigma=\sigma_{ii}/3$ is necessary to describe force
fluctuations.

If we partition the space associating each particle with its
surrounding volume, ${\cal W}_i$ (for instance with a Voronoi
tessellation as it will be done bellow), then the total volume,
$\cal W$, of a system of $N$ particles is given by:
\begin{equation}
{\cal W} = \sum_{i=1}^N {\cal W}_i.
\end{equation}
The ensemble average of the volume function $\cal W$ provides the
volume of the system, $V= \langle {\cal W} \rangle$, in an
analogous way to the average of the Hamiltonian is the energy in
the canonical ensemble of equilibrium statistical mechanics.



The full canonical partition function in the V-F ensemble is the
starting point of the statistical analysis \cite{angoricity}:

\begin{equation}
\mathcal{Z}(X,A) = \int g({\cal W},\Pi) \,\, \exp\Big[
-\frac{\Pi}{A} -\frac{{\cal W}}{X}\Big] \,\,
\Theta_{\rm jam} \,\, d{\cal W} \,\, d\Pi, \label{sw}
\end{equation}
where $g({\cal W},\Pi)$ is the density of states for a given
volume and boundary stress.  Here $\Theta_{\rm jam}$ formally
imposes the jamming restriction and therefore defines the ensemble
of jammed matter. This crucial function will be discussed at
length below. As a minimum requirement it should ensure touching
grains, and obedience to Newton's force laws.

Just as $\partial E/\partial S = T$ is the temperature in
equilibrium systems, the temperature-like variables in granular
systems are the compactivity \cite{edwards2}:
\begin{equation}
X=\frac{\partial V}{\partial S},
\end{equation}
and the angoricity [from the Greek ``$\acute{\alpha}\gamma\chi o
\zeta$'' (ankhos) = stress] \cite{angoricity}
\footnote{Note that in \cite{angoricity}, the angoricity is denoted
$Z$. Here we use $A$ for the isotropic angoricity since $Z$ is
used to denote the mechanical coordination number. In general the
angoricity is a tensor, $A_{ij}=\frac{\partial \Pi_{ij}}{\partial
S}$, here we simplify to the isotropic case.}:
\begin{equation}
A=\frac{\partial \Pi}{\partial S}.
\end{equation}

The compactivity measures the power to compactify while the
angoricity measures the power to stress. These quantities have
remained quite abstract so far, perhaps, this fact being the
primary reason for the great deal of controversy surrounding the
statistical mechanics of jammed matter. In the present paper we
attempt to provide a meaning and interpretation for the
compactivity (the angoricity will be treated in subsequent papers)
by developing equations of states as well interpretations in terms
of a ``compactimeter'' to measure the ``temperature'' of jammed
matter. In Eq. (\ref{sw}), the analogue of the Boltzmann constant
are set to unity for simplicity, implying that the compactivity is
measured in units of volume and the angoricity in units of
boundary stress (stress times volume).

In the limit of vanishing angoricity, $A \rightarrow 0$, the system is
described by the V-ensemble alone. This is the hard sphere limit which
will be the focus of the present work where the following partition
function describes the statistics:
\begin{equation}
\mathcal{Z}^{\rm hard}_{\rm sph}(X) =
\int g({\cal W}) \,\, e^{-{\cal W}/X} \,\, \Theta_{\rm jam} \,\, d{\cal W},
\label{Q}
\end{equation}
where $g({\cal W})$ is the density of states for a given volume
${\cal W}$.


\section{Volume Function}
\label{volume-function}

While it is always possible to measure the total volume of the system,
it is unclear how to partition the space to associate a volume to each
grain.
Thus, the first step to study the V-ensemble is to find the volume
function ${\cal W}_{i}$ associated to each particle that successfully
tiles the system. This is analogous to the additive property of energy
in equilibrium statistical mechanics. This derivation is explained in
Jamming I \cite{jamming1} and is the main theoretical result that
leads to the solution of the partition function for granular
matter. We refer to this paper for details. Here we just state the
main two results, Eqs.  (\ref{vor1}) and (\ref{hamil}), and explain
their significance for the solution of the statistical problem.

Initial attempts at modelling ${\cal W}$ included, (a) a model volume
function under mean-field approximation \cite{edwards2}, (b) the work
of Ball and Blumenfeld \cite{ball,blumenfeld} in 2d and in 3d
\cite{new0}, and (c) simpler versions in terms of the first
coordination shell by Edwards \cite{E8}. These definitions are
problematic: (a) is not given in terms of the contact network, (c) is
not additive, (b) and (c) are proportional to the coordination number
of the balls contrary to expectation (see below).  In Jamming I
\cite{jamming1} we have found an analytical form of the volume
function in any dimensions and
demonstrated that it is the Voronoi volume of the particle.

\begin{figure}
\centering { \hspace{.6cm}\hbox {
\resizebox{5.5cm}{!}{\includegraphics[angle=-90]{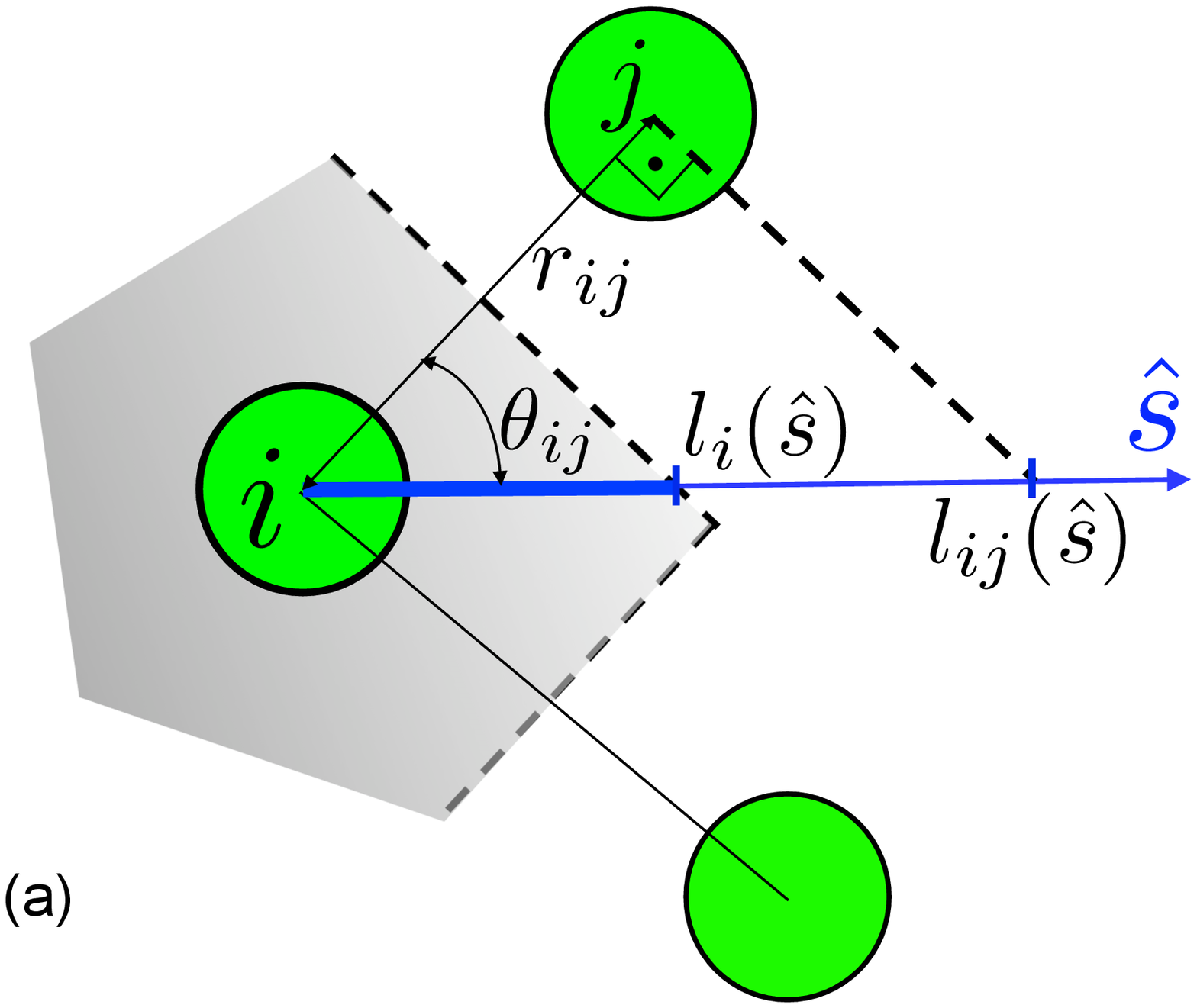}}
}}
\vspace{.5cm}
\caption{ The Voronoi volume is the light grey area (shown in 2d for
  simplicity).  The limit of the Voronoi cell of particle $i$ in the
  direction $\s$ is $l_{i}(\s) = r_{ij}/2 \cos \theta_{ij}$. Then the
  Voronoi volume is proportional to the integration of
  $l_{i}(\hat{s})^3$ over $\s$ as in Eq.  (\ref{vor1}). }
\label{voronoi-figure}
\end{figure}

The definition of a Voronoi cell is a convex polygon whose
interior consists of all points closer to a given particle than to
any other (see Fig. \ref{voronoi-figure}). Its formula in terms of
particle positions for monodisperse spherical packings in 3d is
\cite{jamming1}:

\begin{equation}
  \label{vor1}  {\cal W}_i^{\rm vor} = \frac{1}{3}\left\langle
    \left(\frac{1}{2R}\min_{j}\frac{\rij}{\cos \theta_{ij}}\right)^3
  \right\rangle_s \equiv \langle {\cal W}^s_i\rangle_s,
\end{equation}
where $\vrij$ is the vector from the position of particle $i$ to that
of particle $j$, the
average is over all the directions $\s$ forming an angle $\theta_{ij}$
with $\vrij$ as in Fig. \ref{voronoi-figure}, and $R$ is the radius of
the grain.  ${\cal W}^s_i$ defines the orientational Voronoi volume
which is obtained without the integration over $\s$. This formula has
a simple interpretation depicted in Fig.  \ref{voronoi-figure}.


The Voronoi construction is additive and successfully tiles the total
volume.
Prior to Eq. (\ref{vor1}), there was no analytical formula to
calculate the Voronoi volume in terms of the contact network
$\rij$. Equation (\ref{vor1}) provides such a formula which allows
theoretical analysis in the V-ensemble.  However this microscopic
version is difficult to incorporate into the partition function
since it would necessitate a field theory.  The next step is to
develop a theory of volume fluctuations to coarse grain ${{\cal
W}_{i}^{vor}}$ over a mesoscopic length scale and calculate an
average volume function.  The coarsening reduces the degrees of
freedom to one variable, the coordination number of each grain,
and defines a mesoscopic volume function which is more amenable to
statistical calculations than Eq. (\ref{vor1}). We find a
(reduced) free volume function \cite{jamming1}:
\begin{equation}
  w(z) \equiv \frac {\langle {\cal W}_i^{s} \rangle_i - V_g }{V_g} \approx \frac{2\sqrt{3}}{z},
\label{hamil}
\end{equation}
approximately valid for monodisperse hard spheres with volume $V_g$, where $z$ is
the geometrical coordination number (which is different from the
mechanical coordination number, $Z$, see below). The average is over
all the balls.


The inverse relation with $z$ in Eq. (\ref{hamil}) is in general
agreement with experiments \cite{aste}.  Many experiments have
focused on the analysis of the system volume and single particle
volume fluctuations in granular media
\cite{swinney,aste,cruz,chicago2}.
Of particular importance to the present theory are the recent
advances in X-ray tomography \cite{aste} and confocal microscopy
\cite{brujic,dinsmore} which have revealed the detailed internal
structure of jammed matter allowing for the study of the free
volume per particle. By partitioning the space with Voronoi
diagrams, these studies show that ${{\cal W}_i}^{vor}$ is
distributed with wide tails
\cite{aste,cruz,E8,brujic}. More importantly, the X-ray
tomography experiments with 300,000 monodisperse hard spheres
performed by the Aste group \cite{aste} find that the Voronoi
volumes are inversely proportional, on average, to the
coordination number of the particle, $z$, in reasonable agreement
with the prediction of the volume function Eq. (\ref{hamil}). Such
data is displayed in Fig. 6 in \cite{aste} where the local volume
fraction defined as $\phi_i^{-1} - 1 = {{\cal W}_i}^{vor}/V_g$ is
plotted against the coordination number. From Eq. (\ref{hamil}) we
find:

\begin{equation}
\phi_i = \frac{1} {w + 1} = \frac{z}{z+2 \sqrt{3}}, \label{phi_i}
\end{equation}
which agrees relatively well with the shape of the curves
displayed in Fig. 6 of \cite{aste} for different packing
preparations.  It should be noted that for a more precise
comparison a coarse grained volume fraction and geometrical
coordination number should be considered in Eq. (\ref{phi_i}). A
detailed discussion is provided in Section \ref{bounds}.

The free volume function decreases with $z$ as expected since the more
contacts per grain, the more jammed the particle is and the smaller
the free volume associated with the grain.
The coordination number $z$ in Eq. (\ref{hamil}) can be considered as
a coarse-grained average associated with ``quasiparticles'' with
mesoscopic free volume $w$.
By analogy to the theory of quantum energy spectra, we regard each
state described by Eq. (\ref{hamil}) as an assembly of ``elementary
excitations'' which behave as independent quasiparticles.  As a grain
is jammed in the packing, it interacts with other grains. The role of
this interaction between grains is assumed in the calculation of the
volume function (\ref{hamil}) and it is implicit in the
coarse-graining procedure explained above.  The quasiparticles can be
considered as particles in a self-consistent field of surrounding
jammed matter. In the presence of this field, the volume of the
quasiparticles depends on the surrounding particles as expressed in
Eq. (\ref{vor1}).
The assembly of quasiparticles can be regarded as a set of
non-interacting particles (when the number of elementary excitations
is sufficiently low) and a single particle approximation can be used
to solve the partition function as we will formulate in Section
\ref{phase-diagram}.

While Eq. (\ref{vor1}) is difficult to treat analytically, the
advantage of the mesoscopic Eq.  (\ref{hamil}) is that the partition
function can be solved analytically since $w$ depends on $z$ only,
instead of $\rij$. The key result is the relation between the Voronoi
volume and the coordination number which allow us to incorporate the
volume function into a statistical mechanics approach of jammed hard
spheres, by using the constraint of mechanical stability as we show
below.

\section {\bf Definition of jamming: isostatic conjecture}
\label{definition}

The definition of the constraint function $\Theta_{\rm jam}$ is
intimately related to the proper definition of a jammed state and
should contain the minimum requirement of mechanical equilibrium.


Distinguishing between metastable and mechanically stable packings
that define the jammed state through the $\Theta_{\rm jam}$ function
remains a problem under debate, related to the more fundamental
question of whether or not a jammed packing is well-defined
\cite{torquato}.  In practice, it is widely believed that the
isostatic condition is necessary for a jammed disordered packing
following the Alexander conjecture
\cite{alexander,moukarzel,edwards-grinev} which was tested in several
works \cite{makse,ohern,silbert,kertesz}


It is well known that mechanical equilibrium imposes an average
coordination number larger or equal than a minimum coordination where
the number of force variables equals the number of force and torque
balance equations \cite{alexander,edwards-grinev,moukarzel}. The so-called
isostatic condition.


In the case of frictionless spherical particles the isostatic
condition is $Z=2d=6$ (in 3d), while the coordination in the case
of infinitely rough particles (with interparticle friction
coefficient $\mu\to \infty$) is $Z=d+1=4$, where $d$ is the
dimension of the system.  Numerical simulations and theoretical
work suggest that at the jamming transition the system becomes
exactly isostatic
\cite{ohern,ohernprime,makse,vanhecke1,edwards-grinev,kertesz,moukarzel}. But no
rigorous proof of this statement exists. In the following, we will
use this isostatic conjecture to define the ensemble of jammed
matter.

While isostaticity holds for perfectly smooth and infinitely rough
grains, the main problem is to extend it to finite frictional systems.
For finite friction, the Coulomb condition takes the form of an
inequality between the normal interparticle contact force $F_n$ and
the tangential one $F_t$: $F_t\le \mu F_n$ and therefore no trivial
solution to the minimum number of contacts can be obtained.

The problem can be understood as an optimization of an outcome
based on a set of constraints, i.e., minimizing a Hamiltonian of a
system
over a convex polyhedron specified by linear and non-negativity
constraints.  The isostatic condition can be augmented to indicate
the number of extra equations for contacts satisfying the Coulomb
condition, which is analogous to the number of redundant
constraints in Maxwell constraint counting of rigidity
percolation.  This suggests that, for a finite value of $\mu$, the
original nonlinear problem can be mapped to a linear equation
problem if we know how many extra equations should be added. This
problem is treated with more details in a future paper.

In the following, we present a rough approximation suggesting that the
coordination number could be related to friction.  We consider the
following argument.
Consider a set of spherical particles interacting via normal and
tangential contact forces. These can be the standard Hertz and
Mindlin/Coulomb forces of contact mechanics
\cite{landau,mindlin,zhang}, respectively, see Section \ref{md}.  We
set $N$: number of particles, $N_n$: number of unknown normal forces,
$N_t$: number of unknown tangential forces, $E_f$: number of force
balance equations, $E_t$: number of torque balance equations, $Z = 2
M/N$: average coordination number of the packing, where $M$ is the
total number of contacts, $f_1(\mu)$: undetermined function of the
friction coefficient $\mu$ such that $1-f_1(\mu)$ is the fraction of
spheres that can rotate freely
($f_1(0)=0$ and $f_1(\infty)=1$), and $f_2(\mu)$: undetermined
function of $\mu$ indicating the ratio of contacts satisfying $F_t<\mu
F_n$,
which satisfies $f_2(0)=0$ and $f_2(\infty)=1$.

A packing is isostatic when

\begin{equation}
  N_n+N_t=E_f+E_t.
\label{nt}
\end{equation}
The average coordination number at the isostatic point is then (see
Table \ref{table1}):
\begin{equation}
  Z(\mu)=2d \frac{1+1/2 (d-1) f_1(\mu)}{1+ (d-1) f_2(\mu)},
 \label{Eq_Zc}
\end{equation}
reducing to the known $Z=2d$ for frictionless particles and $Z=d+1$
for infinitely rough particles.

\bigskip
\begin{table}
  \caption{\label{table1}
    Number of constraints and variables determining the isostatic condition
    for different systems of spherical particles.}
\vspace{.5cm}
\begin{tabular}{|c|c|c|c|c|}
  \hline
  Friction&$N_n$&$N_t$&$E_f$&$E_t$\\
  \hline
  $\mu=0$&$\frac{1}{2}NZ$&0&$dN$&0\\
  \hline
  $\mu$ finite&$\frac{1}{2}NZ$&$\frac{1}{2}(d-1)NZf_2(\mu)$&$dN$&$\frac{1}{2}d(d-1)Nf_1(\mu)$\\
  \hline
  $\mu=\infty$&$\frac{1}{2}NZ$&$\frac{1}{2}(d-1)NZ$&$dN$&$\frac{1}{2}d(d-1)N$\\
  \hline
\end{tabular}
\end{table}

In what follows, we use the numerical fact that interpolating
between the two isostatic limits, there exist packings of finite
$\mu$ with the coordination number smoothly varying between
$Z(\mu=0)=6$ and $Z(\mu\to \infty)\to 4$ \cite{silbert}.
Numerical simulations with packings described in Section
\ref{simulations} corroborate this result and further show that the
$Z$ vs $\mu$ dependence is independent of the preparation protocol as
obtained in our simulations (see Fig. \ref{z}). This result then
generalizes the isostatic conditions from $\mu=0$ and $\mu = \infty$
to finite $\mu$.

It should be noted that we cannot rule out that other preparation
protocols could give rise to other dependence of $Z$ on
$\mu$. However, we will see that the obtained phase diagram is given
in terms of $Z$; the main prediction of the theory would still be
valid irrespective of the particular dependence $Z(\mu)$.  That is,
the theory does not assume anything about the relation between the
interparticle friction and $Z$.


It is worth noting that other attempts to define the jammed state
have been developed. A rigorous attempt is that of Torquato {\it
et al.} who propose three categories of jamming \cite{salvatore1}:
locally jammed, collectively jammed and strictly jammed based on
geometrical constraints.
However, the definition of \cite{salvatore1} is based purely on
geometrical considerations and therefore only valid for
frictionless particles. Thus it is not suitable for granular
materials with inter-particle frictional tangential forces; their
configurational space is influenced by the mechanics of normal and
shear forces. Other approaches to define a jammed state based on
the potential energy landscape \cite{ohern} fail for granular
materials too since such a potential does not exist for frictional
grains due to their inherent path-dependency.  Thus, the
definition of jammed state for granular materials must consider
interparticle normal and tangential contact forces beyond
geometry. In the companion paper \cite{jamming3} we elaborate on
this problem.

\subsection{
Geometrical and mechanical coordination number}
\label{geometrical}

We have acknowledged a difference between the geometrical
coordination number $z$ in Eq. (\ref{hamil}) and the mechanical
coordination number $Z$ which counts only the contacts with
non-zero forces.  Below we discuss the bounds of $z$.

Since some geometrical contacts may carry no force, then we have:

\begin{equation}
  Z\le z. \label{coord_low}
\end{equation}
To show this, imagine a packing of infinitely rough ($\mu\to \infty$)
spheres with volume fraction close to $0.64$. There must be $z=6$
nearest neighbors around each particle on the average. However, the
mechanical balance law requires only $Z=4$ contacts per particle on
average, implying that $2$ contacts have zero force and do not
contribute to the contact force network.

Such a situation is possible as shown in Fig. \ref{possible}: starting
with the contact network of an isostatic packing of frictionless
spheres having $Z=6$ and all contacts carrying forces (then $z=6$ also
as shown in Fig. \ref{possible}a), we simply allow the existence of
tangential forces between the particles and switch the friction
coefficient to infinity. Subsequently, we solve the force and torque
balance equations again for this modified packing of infinitely rough
spheres but same geometrical network, as shown in Fig. \ref{possible}b
[Notice that the shear force is composed of an elastic Mindlin
component plus the Coulomb condition determined by $\mu$, see Section
\ref{md} for details.  Thus when $\mu \to \infty$, the elastic Mindlin
component still remains].

The resulting packing is mechanically stable and is obtained by
setting to zero the forces of two contacts per ball, on average, to
satisfy the new force and torque balance condition for the additional
tangential force at the contact.  Such a solution is guaranteed to
exist due to the isostatic condition: at $Z=4$ the number of equations
equals the number of force variables.  Despite mechanical equilibrium,
giving $Z=4$, there are still $z=6$ geometrical contacts contributing
to the volume function.

Therefore, we identify two types of coordination number: the
geometrical coordination number, $z$, contributing to the volume
function and the mechanical coordination number, $Z$, measuring
the contacts that carry forces only.  This distinction is crucial
to understand the sum over the states and the bounds in the
partition function as explained below.

These ideas are corroborated by numerical simulations in Section
\ref{bounds}, below.  The packings along the vertical RCP line
found in the simulations (see Fig. \ref{phaseb}) have
approximately the same geometrical coordination number, $z\approx
6$.  However, they differ in mechanical coordination number, going
from the frictionless point $Z=6$ to $Z \approx 4$ as the friction
coefficient is increased to infinity.



We have established a lower bound of the geometrical coordination in
Eq. (\ref{coord_low}). The upper bound arises from considering the
constraints in the positions of the rigid hard spheres.  For hard
spheres, the $Nd$ positions of the particles are constrained by the
$Nz/2$ geometrical constraints, $|\vec{r}_{ij}| = 2R$, of rigidity.
Thus, the number of contacts satisfies $Nz/2\le Nd$, and $z$ is
bounded by:

\begin{equation}
  z\le 2d. \label{coord1}
\end{equation}
Notice that this upper bound applies to the geometrical coordination,
$z$ and not to the mechanical one, $Z$, and it is valid for any system
irrespective of the friction coefficient, from $\mu = 0 \to \infty$.

Furthermore, in relation with the discussion of frictionless
isostaticity, it is believed that above $2d$, the system is partially
crystallized. To increase the coordination number above 6, it is
necessary to create partial crystallization in the packing, up to the
point of full order of the FCC lattice with coordination number 12.
Thus, by defining the upper bound at the frictionless isostatic limit
we also exclude from the ensemble the partially crystalline packings.
This is an important point, akin to mathematical tricks employed in
replica approaches to glasses \cite{parisi}.

In conclusion, the mechanical coordination number, $Z$, ranges from 4
to 6 as a function of $\mu$, and provides a lower bound to the
geometrical coordination number, while the upper bound is $2d$.  A
granular system is specified by the interparticle friction which
determines the average mechanical coordination at which the system is
equilibrated, $Z(\mu)$. The possible microstates in the ensemble
available for this system follow a Boltzmann distribution
Eq. (\ref{Q}) for states satisfying the following bounds:
\begin{equation}
\label{bounds-eq}
Z(\mu) \le z \le 2d = 6.
\end{equation}

\section{Density of states}
\label{dos}

According to the statistical mechanics of jammed matter, the
volume partition function $\mathcal{Z}$ is defined by Eq.
(\ref{Q}). In the quasiparticle approximation we can write:

\begin{equation}
\begin{split}
  \mathcal{Z}^{\rm hard}_{\rm sph}(X) =\int\ldots\int
  g\Big(\sum w(z_i)\Big) \,\, e^{-\sum w(z_i)/X} \times \\
  \times \,\, \Theta_{\rm jam} \,\, \prod_i^N dw(z_i).
\end{split}
\label{Q2}
\end{equation}

Considering $N$ non-interacting quasiparticles with free volume
$w(z)$, the partition function can be written as:

\begin{equation}
\label{partw} \mathcal{Z}_{\rm sph}^{\rm hard} = \Big(
\int g(w) \,\, e^{-w/X} \,\,  \Theta_{\rm jam} \,\,dw \Big)^N.
\end{equation}
Here, $g(w)$ is the density of states for a given quasiparticle free
volume.

Since the mesoscopic $w$ is directly related to $z$ through
Eq. (\ref{hamil}), we change variables to the geometrical coordination
number in the partition function.  The density of states for a single
quasiparticle, $g(w)$, is:
\begin{equation}
g(w) = \int_{Z}^6 P(w|z) g(z) dz,
\end{equation}
where $P(w|z)$ is the conditional probability of a free
volume $w$ for a given $z$, and $g(z)$ is the density of
states for a given $z$. Here, we have used the bounds in Eq.
(\ref{bounds-eq}).

The next step in the derivation is the calculation of $g(z)$ which
is developed in three steps:

First, we consider that the packing of hard spheres is in a jammed
configuration where there can be no collective motion of any
contacting subset of particles leading to unjamming when including
the normal and tangential forces between the particles. This
definition is an extended version of the collectively jammed
category proposed by Torquato {\it et al.} \cite{salvatore1} that
goes beyond the merely locally jammed configuration of packings,
unstable to the motion of a single particle. While the degrees of
freedom are continuous, the fact that the packing is collectively
jammed implies that the jammed configurations in the volume space
are not continuous. Otherwise there would be a continuous
transformation in the position space that would unjam the system
contradicting the fact that the packing is collectively jammed.

Thus, we consider that the configuration space of jammed matter is
discrete since we cannot change one configuration to another in a
continuous way. Similar consideration of discreteness has been studies
in \cite{salvatore1}. Notice that the volume landscape could be
continuous since we could change the volume as well, or, in the case
of soft particle, we can deform them. Additionally, in the case of
frictionless packings of soft particle, the energy of deformation is
well-defined, and we can define the collectively jammed configuration
as a minimum in the energy with definitive positive Hessian (or a zero
order saddle) like in \cite{ohern}. In this case, there could be no
continuous transformation of the particle coordinates that brings one
jammed state to the next, unless we deform the particles.  Thus, the
space is discrete in this case as well.

Second, we call the dimension per particle of the configuration space
as $\mathcal{D}$ and consider that the distance between two jammed
configurations is not broadly distributed (meaning that the average
distance is well-defined). We call the typical (average) distance
between configurations in the volume space as $h_z$, and therefore the
number of configurations per particle is proportional to
$1/(h_z)^{\mathcal{D}}$.  The constant $h_z$
plays the role of the Planck's constant, $h$, in quantum
mechanics.

Third, we add $z$ constraints per particle due to the fact that
the particle is jammed by $z$ contacts. Thus, there are $Nz$
position constrains ($|\vec{r}_{ij}|=2R$ ) for a jammed state of
hard spheres as compared to the unjammed ``gas'' state. Therefore,
the number of degrees of freedom is reduced to $\mathcal{D}-z$,
and the number of configurations is then
$1/(h_z)^{\mathcal{D}-z}$. Since the term $1/(h_z)^{\mathcal{D}}$
is a constant, it will not influence the average of the
observables in the partition function (although it changes the
value of the entropy, see Jamming III \cite{jamming3}).
Therefore, the density of states $g(z)$ is assumed to have an
exponential form:
\begin{equation}
g(z) = (h_z)^{z} = e^{-z/z_c},
\label{gz}
\end{equation}
with $z_c^{-1}\equiv \ln(1/h_z)$. Physically, we expect $h_z\ll 1$,
then $g(z)$ is an exponentially rapid decreasing function with
$z$. The exact value of $h_z$ can be determined from a
fitting of the theoretical values to the simulation data. The most populated state is the highest volume at $z=4$ while the
least populated state is the ground state at $z=6$.  A constant could
be added in (\ref{gz}) but it has no consequence for the average of
the observables in the partition function.  However, it affect the
value of the entropy. For the calculation of the entropy we consider
that there is a single mesoscopic ground state and use $g(z) =
e^{-(z-2d)/z_c}$ \cite{jamming3}.

We see that the negative of the geometrical coordination number,
$-z$, plays the role of a
number of degrees of
freedom for a packing, due to the extra position constrains of the
contacting particles.  The coordination $z$ can be then considered
as the number of degrees of ``frozen'' per particle.  Another way
to understand Eq. (\ref{gz}) is the following.  In the case of a
continuous phase space of configurations we would obtain
$\frac{g(z+1)}{g(z)} = 0$. However, since the space of volume
configurations is discrete as discussed above, the ratio
$\frac{g(z+1)}{g(z)} \sim h_z$. This implies again Eq. (\ref{gz}).

The conditional probability $P(w|z)$ depends on the $w$ function,
$w = \frac{2 \sqrt{3}}{z}$. The average is taken over a certain
mesoscopic length scale since the volume of a particle depends on the
positions of the particles surrounding it. Practically, such length
scale is approximately of several particle diameters.  $w$ is a
coarse-grained volume and independent of the microscopic partition of
the particles, implying:
\begin{equation} \label{Pwz}
P(w|z) = \delta (w-2\sqrt{3}/z).
\end{equation}
The meaning of Eq. (\ref{Pwz}) is that we neglect the fluctuations in
the coordination number due to the coarse graining procedure. A more
general ensemble can be considered where the fluctuations in the
geometrical coordination number are taken into account.  We have
extended our calculations to consider fluctuations in $z$ and find a
similar phase diagram as predicted by the present partition
function. This generalized z-ensemble will be treated in a future
paper, where the boundaries of the phase diagram are regarded as a
second-order phase transition.  The generalized z-ensemble allows for
the calculation of the probability distribution of the coordination
number, beyond the assumption of the delta function distribution in
Eq. (\ref{Pwz}).  The resulting prediction of $P(z)$ is in general
agreement with simulations (see \cite{jamming4}).

Substituting Eq. (\ref{Pwz}) and Eq. (\ref{gz}) into Eq.
(\ref{partw}), we find the isostatic partition function which is used
in the remaining of this study:

\begin{equation} \label{pf1}
\mathcal{Z}_{\rm iso}(X,Z) = \int_{Z}^6 (h_z)^{z}
\exp\left(-\frac{2 \sqrt{3}}{z X}\right )
dz.
\end{equation}

\section{Phase diagram}
\label{phase-diagram}

Next, we obtain the equations of state to define the phase diagram of
jammed matter by solving the partition function.
From Eq. (\ref{pf1}), we calculate the ensemble average volume
fraction $\phi= (w + 1)^{-1} = z/(z+2\sqrt{3})$ as:
\begin{eqnarray}
\begin{split}
&\phi(X,Z)=&\\
&\frac{1}{\mathcal{Z}_{\rm iso}(X,Z)} \int_{Z}^6
\frac{z}{z+2\sqrt{3}} \exp\left(-\frac{2 \sqrt{3}}{z X}+z \ln
h_z\right)dz.&
\end{split}
\label{phi1}
\end{eqnarray}

We start by investigating the limiting cases of zero and infinite
compactivity.

(a) In the limit of vanishing compactivity ($X\to 0$),
only the minimum volume or ground state at
$z=6$ contributes to the partition function. Then
we obtain the ground state of jammed matter with a density:

\begin{equation}
  \phi_{\rm RCP} = \phi(X=0,Z) = \frac{6}{6+2\sqrt{3}} \approx 0.634,
  \,\, \,\, \,\, Z(\mu)\in[4,6].
\label{state1}
\end{equation}
The meaning of the subscript RCP in (\ref{state1}) will become clear
below.



(b) In the limit of infinite compactivity ($X \to \infty$), the
Boltzmann factor $\exp[-2 \sqrt{3}/(z X)] \to 1$, and the average in
(\ref{phi1}) is taken over all the states with equal probability. We
obtain:

\begin{equation}
\begin{split}
 \phi_{\rm RLP}(Z) = &\phi(X\to \infty,Z) = \\
= &
\frac{1}{\mathcal{Z}_{\rm iso}(\infty,Z)}\int_Z^6
  \frac{z}{z+2\sqrt{3}}\exp\left(z \ln h_z\right )dz.
\end{split}
\label{phi4}
\end{equation}

The constant $h_z$ determines the minimum volume in the phase
space.  We expect $h_z\ll 1$, such that the exponential in Eq.
(\ref{phi4}) decays rapidly. Then the leading contribution to Eq.
(\ref{phi4}) is from the highest volume at $z=Z$ and therefore:
\begin{equation}
\begin{split}
  \phi_{\rm RLP}(Z) & \approx \frac{Z}{Z+2\sqrt{3}}, \,\, \,\, \,\,
  Z(\mu)\in[4,6].
\end{split}
\label{state2}
\end{equation}

This dependence of the volume fraction on $Z$ suggests using the
$(\phi,Z)$ plane to define the phase diagram of jammed matter as
plotted in Fig. \ref{phasea}.  The equations of state
(\ref{state1}) and (\ref{state2}) are plotted in the $(\phi,Z)$
plane in Fig. \ref{phasea} providing two limits of the phase
diagram.  Since the mechanical coordination number is limited by
$4\le Z \le 6$ we have two more horizontal limits: The phase space
is delimited from below by the minimum coordination $Z=4$ for
infinitely rough grains, denoted the granular-line or G-line in
Fig. \ref{phasea}.

All mechanically stable disordered jammed packings lie within the
confining limits of the phase diagram (indicated by the yellow zone in
Fig. \ref{phasea}), while
the grey shaded area in Fig. \ref{phasea} indicates the forbidden
zone.  For example, a packing of frictional hard spheres with
$Z=5$ (corresponding to a granular material with interparticle
friction $\mu\approx 0.2$ according to Fig. \ref{z}) cannot be
equilibrated at volume fractions below $\phi< \phi_{\rm RLP}(Z=5)
= 5/(5+2\sqrt{3}) = 0.591$ or above $\phi>\phi_{\rm RCP}=0.634$.
It is worth noting that particular packings can exist in the
forbidden zone; our contention is that they are zero measure and
therefore have zero probability of occurrence at the ensemble
level.

These results provide a statistical interpretation of the RLP and RCP limits:


{\bf {\it (i)} The RCP limit.--} Stemming from the
statistical mechanics approach, the RCP limit arises as the result
of the
relation (\ref{state1}), which gives the maximum volume fraction of
disordered packings under the mesoscopic framework.  To the right of
the RCP-line, packings exist only with some degree of order (for
instance with crystalline regions).  The prediction,
\begin{equation}
  \phi_{\rm RCP}=\frac{6}{6+2\sqrt{3}}\approx 0.634,
\label{RCP_state}
\end{equation}
is valid for all friction coefficients and approximates the
experimental and numerical estimations
\cite{anonymous,scott,finney,berryman} which find a close packing
limit independent of friction in a narrow range around 0.64.

Beyond the fact that 63-64\% is commonly quoted as RCP for
monodisperse hard spheres, we present a physical interpretation
of that value as the ground state of frictional hard spheres
characterized by a given interparticle friction coefficient.  In this
representation, as $\mu$ varies from 0 to $\infty$ and $Z$ decreases
from $6$ to $4$, the state of RCP changes accordingly while its volume
fraction remains the same, given by Eq. (\ref{RCP_state}). The present
approach has led to an
unexpected number of states that all lie in the RCP line from $Z=6$ to
$Z=4$ as depicted in Fig. \ref{phasea}, suggesting that RCP is not a
unique point in the phase diagram.

An important prediction is that for frictionless systems there is
only one possible state at $Z=6$. It is important to note that
there is one state only at the mesoscopic level used in the
theory. However, for a single mesoscopic state, we expect many
microstates, which are averaged out in the mesoscopic theory of
the volume function. Thus, there could be more jammed states
surrounding the frictionless point in the phase diagram. However,
we expect that these states are clustered in a narrow region
around the frictionless point.  To access these microstates it
would require different preparation protocols, analogous to the
dependence of the glass transition temperature on cooling rates in
glasses \cite{skoge,parisi}.

It is interesting to note that replica approaches to the jammed states
\cite{parisi} predict many jammed isostatic states with different
volume fractions for frictionless hard spheres.  The first question is
whether these states in the force/energy ensemble are the same as the
volume ensemble states that we treat in our approach. It should be
noted that jamming in \cite{skoge,parisi} is obtained when the
particles are rattling infinitely fast in their cages, in the limit of
infinite number of collisions per unit time. That is when the dynamic
pressure related to the momentum of the particles diverges. Second,
from our point of view, an investigation of the fluctuations of the
microstates as well as more general ensembles allowing for
fluctuations in the coordination number could be considered. These
studies, may reveal whether the frictionless point is unique or
not. This point is investigated further in Section \ref{micro_states}.

{\bf {\it (ii)} The RLP limit.---} Equation of state
(\ref{state2}) provides the lowest volume fraction for a given $Z$
and represents a statistical interpretation of the RLP limit depicted by the RLP-line in
Fig. \ref{phasea}. We predict that to the left of this line,
packings are not mechanically stable or they are experimentally
irreversible as discussed in \cite{chicago2,swinney,sdr}.

A review of the literature indicates that there is no general
consensus on the value of RLP as different estimations have been
reported ranging from 0.55 to 0.60 \cite{scott,onoda,berryman},
proposing that there is no clear definition of RLP limit. The
phase diagram proposes a solution to this problem. Following the
infinite compactivity RLP-line, the volume fraction of the RLP
decreases with increasing friction from the frictionless point
$(\phi,Z)=(0.634,6)$, towards the limit of infinitely rough hard
spheres, $Z\rightarrow4$. Indeed, experiments \cite{scott}
indicate that lower volume fractions are achieved for larger
coefficient of friction.  We predict the lowest volume fraction in
the limit: $\mu \to \infty$, $X\to \infty$ and $Z\to 4$ (and
$h_z\to 0$) at

\begin{equation}
\phi_{\rm RLP}^{\rm min} = \frac{4}{4+2\sqrt{3}} \approx 0.536.
\label{rlp_min}
\end{equation}
Even though this is a theoretical limit, our results indicate that for
$\mu > 1$ this limit can be approximately achieved.

The finding of a random loose packing bound is an interesting
prediction of the present theory. The RLP limit has not been well
investigated experimentally, and so far it was not certain whether
this limit can or cannot be reached in real systems.
The lowest stable volume fraction ever reported, $0.550\pm0.006$,
obtained by Onoda and and Liniger \cite{onoda} as the limit of
vanishing gravity for spherical glass beads, is not far from the
present prediction.

The intersections of the RCP, RLP and the G-line identify three
interesting points in the $(\phi,Z(\mu))$ plane:

(a) The frictionless point $\mu=0$, denoted J-point in \cite{ohern},
at $$J \equiv (\phi_{\rm RCP},Z(0)) = (0.634,6),$$ corresponds to a
system of compressed emulsions in the limit of small osmotic pressure
as measured by J. Bruji\'c \cite{E21}.

(b) The lowest coordination number $Z=4$ plotted as the G-line
defines two associated points from the lowest volume fraction of loose
packings at infinite compactivity, L-point, $$L \equiv (\phi^{\rm
  min}_{\rm RLP},Z(\infty)) = (0.536,4),$$ to the zero compactivity
state of close packing, C-point, $$C\equiv (\phi_{\rm RCP},Z(\infty))
= (0.634,4).$$ The full JCL triangle defines the isostatic plane where
the frictional hard sphere packings reside.

{\bf {\it (iii)} Intermediate isocompactivity states.---} For
finite $X$, Eq. (\ref{phi1}) can be solved numerically.  For each
$X$, the function $\phi(X,Z)$ can be obtained and is plotted as
each isocompactivity color line in Fig. \ref{phasea}.  Between the
two limits Eqs. (\ref{state1}) and (\ref{state2}), there are
packings inside the yellow zone in Fig. \ref{phasea} with finite
compactivity, $0<X<\infty$. Since $X$ controls the probability of
each state, like in condensed matter through a Boltzmann-like
factor in Eq.  (\ref{Q}), it characterizes the number of possible
ways to rearrange a packing having a given volume and entropy,
$S$.  Thus, the limit of the most compact and least compact stable
arrangements correspond to $X\to0$ and $X\to \infty$,
respectively. Between these limits, the compactivity determines
the volume fraction from RCP to RLP.

{\bf Dependence on $h_z$ and negative compactivity.--}

Of interest is the dependence of our prediction on the ``Planck
constant'' of jammed matter, $h_{z}$, that determines the minimum
volume in the phase space.  The equation of state (\ref{state2}) and
the prediction of the minimum RLP, Eq.  (\ref{rlp_min}) have been
obtained by considering $h_z\to 0$ but still nonzero. Thus, when
$X\to\infty$, the only state contributing to the volume partition
function is the most populated at $z=Z(\mu)$.

The approximation $h_z\ll 1$ is a sensible one, since the
discretization of the space is supposed to be very small. However,
this constant remains a fitting parameter of the theory. Indeed in the
simulations we will use a value of $h_z=e^{-100}$ in order to fit the
theoretical values with the numerical ones for finite
compactivity. This extremely small constant shifts the value of the
minimum RLP a little bit to the right of the phase diagram from the
prediction of Eq.  (\ref{rlp_min}). Indeed, when we plot the phase
diagram of Fig.  \ref{phasea} with a $h_z=e^{-100}$ a slightly larger
value of $\phi_{\rm RLP}^{\rm min}$ is obtained as seen in
Fig. \ref{phasea}. In the unphysical limit of $h_z\to 1$ we would
obtain a minimum RLP value of $(\phi_{\rm RLP}^{\rm min} + \phi_{\rm
  RCP})/2$, although as said, this would correspond to an unphysical
situation.

However, it should be noted that this argument depends on the assumed
exponential form of the density of states. Since at this point we do
not have the exact form for the density of states, the results could
change if other more accurate density is found to be valid.  On the
other hand the prediction of the ground state at RCP,
Eq. (\ref{state1}), remains unaffected by the density of state or the
value of $h_z$.

It is interesting to note that we can extend the compactivity to
negative values and study the range $X:0^-\to -\infty$. Indeed, for
any value of $h_z$, in the limit $X\to 0^-$ we obtain the lowest
volume fraction of the prediction of Eq.  (\ref{state2}). Thus,
properly speaking, the minimum value of RLP is defined in the limit of
negative zero compactivity, and this limit is independent of the value
of $h_z$. The entropy has an interesting behavior in this regime which
will be discussed later.

It is important to realize that the region from $X:0^-\to\infty$
disappears when $h_z\to 0$, thus the only meaningful limit is that of
$X\to+\infty$ (which equals both the limits $X\to-\infty$ and
$X\to0^-$ when $h_z\to 0$).  Thus, for any practical purpose,
Eq. (\ref{rlp_min}) can be considered to be the lowest possible
density predicted by the theory. However, notice that the shape of the
iso-compactivity curves depends slightly on the value of $h_z$.

Since we always expect $h_z\ll 1$, it may not be necessary to use the
negative compactivity states to describe RLP.
We leave this interesting observation for further
investigations where the entropy of jamming is treated in more details
in \cite{jamming3}.



\subsection{Volume landscape of jammed matter}

Equation (\ref{hamil}) plays the role of the Hamiltonian of the jammed
system and the jammed configurations can be considered as the minima
of (\ref{hamil}).
Inspired by the physics of glasses and supercooled liquids, we
imagine a "volume landscape", analogous to the energy landscape in
glasses \cite{stillinger}, as a pictorial representation to
describe the states of jammed matter.  Each mesoscopic jammed
state (determined by the positions of the particles, denoted $\vec
r_i$, and its corresponding volume $w$) is depicted as a point in
Fig. \ref{landscape}. At the mesoscopic level, the volume
landscape has different levels of constant $z$, analogous to
energy levels in Hamiltonian systems.

The lowest volume corresponds to the FCC/HCP structure (with kissing
number $z=12$), as conjectured by Kepler \cite{hales}.  Other lattice
packings, such as the cubic lattice and tetrahedron lattice, have
higher volume levels in this representation. Beyond these ordered
states, the ensemble of disordered packings is identified within the
yellow area in Fig. \ref{landscape}a, corresponding to a system with
infinite friction.  In this case the partition function is integrated
from $z=4$ to 6, thus all the states are sampled in the configuration
average as indicated by the arrow in Fig. \ref{landscape}a.  When
$X\to\infty$, all the states are sampled with equal probability, and,
when $X\to 0$, the ground state is the most probable. As the
compactivity is varied, the states along the G-line in the phase
diagram result.
For $\mu\to \infty$, the maximum volume, $w(z=4) = \sqrt{3}/2$ is
attained for $z=4$ when $\mu \to \infty$ in analogy with the high
energy states in a classical system.

As we set the friction coefficient to a finite value, the
available states in the volume are less, since the integration in
the partition function is in the region $Z(\mu)\le z \le 6$. This
is indicated in Fig. \ref{landscape}b for a generic $Z(\mu)$.
Finally when the friction vanishes, we obtain only the ground
state $z=6$ as indicated in Fig. \ref{landscape}c.

For any value of $\mu$, the lowest state is always at $z=6$.  This
corresponds to the states exemplified by the RCP-line in the phase
diagram, all of them with a geometrical coordination $z=6$.  Equation
(\ref{state1}) indicates that the RCP corresponds to the ground state
of disordered jammed matter for a given friction which determines $Z$,
while
the RLP states are achieved for higher volume levels as indicated
in Fig. \ref{landscape}.  An important conclusion is the
following:  The states along the RCP line all have $z = 6$,
independent of $Z$ (and $h_z$) which ranges from 4 to 6 as a
function of the friction coefficient $\mu$.  The states along the
RLP line all have $z = Z$ (when $h_z \to 0$).  These predictions
find good agreement with the numerically generated packings in
Section \ref{bounds}.

%

At fixed volume, the jammed states are separated by barriers of
deformation energy, as depicted in Fig. \ref{landscape}d.  These
barriers can be understood as follows: so far we have treated the
case of hard spheres considered as soft spheres (interacting via
soft-potential such as Hertz-Mindlin forces) in the limit of
vanishing deformation or infinite shear modulus. Indeed, the way
to consider forces in granular materials is by considering the
small tiny deformation at the contact points and a given force law
\cite{zhang}.
In a sense, deformable particles are needed when discussing realistic
jammed states especially when considering the problem of sound
propagation and elastic behavior \cite{elastic,mgsj-pre}. In the case
of deformable particles the third axis in Fig. \ref{landscape}d
corresponds to the energy of deformation or the work done to go from
one configuration to the next. This energy is not uniquely defined in
terms of the particle coordinates; it depends on the path taken from
one jammed state to the next. Thus, we emphasize that the energy in
Fig. \ref{landscape} is path-dependent. The only point where it
becomes independent of the path is in the frictionless point. Besides
this, the volume landscape in the isostatic plane
Figs. \ref{landscape}a-c is well defined and independent of the energy
barriers and path dependent issues.

It is important to note that the basins in Fig. \ref{landscape} are
not single states, but represent many microscopic states with
different degrees of freedom $\vec{r_i}$, parameterized by a common
value of $z$ with a density of states $g(z)$.  The basins represent
single states only at the mesoscopic level providing a mesoscopic view
of the landscape of jammed states.  This is an important distinction
arising from the fact that the states defined by $w(z)=2\sqrt{3}/z$,
Eq. (\ref{hamil}), are coarse grained from the microscopic states
defined by the microscopic Voronoi volume Eq. (\ref{vor1}) in the
mesoscopic calculations leading to (\ref{hamil}) as discussed in
Jamming I \cite{jamming1}. This fact has important implications for
the present predictions which will be discussed in Section
\ref{micro_states}.  The advantage of the volume landscape picture is
that it allows visualization of the corresponding average over
configurations that give rise to the macroscopic observables of the
jammed states.

\subsection{Equations of state}

Further statistical characterization of the jammed structures can be
obtained through the calculation of the equations of state in the
three-dimensional space $(X,\phi,S)$, with $S$ the entropy, as seen in
Fig.  \ref{entropy}.

The entropy density, $s = \frac {S}{N}$, is obtained as:

\begin{equation}
s(X,Z) = \langle w \rangle /X + \ln {\mathcal Z}_{\rm iso}(X,Z)
\end{equation}

This equation is obtained in analogy with equilibrium statistical
mechanics and it is analogous to the definition of free energy: $F=E -
T S$ where $F=- T \ln {\mathcal Z}$ is the free energy.
We replace $T\rightarrow X$, $E\rightarrow \left<w\right>$.
Therefore, $F = E - TS$ or $S = (E-F)/T = E/T + \ln {\mathcal Z} $ is
now $s(X,Z) = \left<w\right>/X + \ln {\mathcal Z}_{\rm iso}(X,Z)$,
which is plotted as the equation of state in Fig.  \ref{entropy}.


Each curve in the figure corresponds to a system with a different
$Z(\mu)$.  The projections $S(X)$ and $S(\phi)$ in Fig.
\ref{entropy} characterize the nature of randomness in the
packings.  When comparing all the packings, the maximum entropy is
at $\phi_{\rm RLP}^{\rm min}$ and $X\to \infty$ while the entropy
is minimum for $\phi_{\rm RCP}$ at $X\to 0$. Following the G-line
in the phase diagram we obtain the entropy for infinitely rough
spheres showing a larger entropy for the RLP than the RCP. The
same conclusion is obtained for the other packings at finite
friction ($4<Z(\mu)<6$). We conclude that the RLP states are more
disordered than the RCP states. Approaching the frictionless
$J$-point at $Z=6$ the entropy vanishes.  More precisely, it
vanishes for a slightly smaller $\phi$ than $\phi_{\rm RCP}$ of
the order $h_{z}$.  Strictly speaking, the entropy diverges to
$-\infty$ at $\phi_{\rm RCP}$ as $S \rightarrow \ln X$ for any
value of $Z$, in analogy with the classical equation of state,
when we approach RCP to distances smaller than $h_z$.  However,
this is an unphysical limit, as it would be like considering
distances in phase space smaller than the Planck constant.

It is commonly believed that the RCP limit corresponds to a state
with the highest number of configurations and therefore the
highest entropy.  However, here we show that the states with a
higher compactivity have a higher entropy, corresponding to looser
packings. Within a statistical mechanics framework of jammed
matter, this result is a natural consequence and gives support to
such an underlying statistical picture. A more detailed study of
the entropy is performed in Jamming III \cite{jamming3}.

The interpretation of the RCP as the ground state, $X \rightarrow 0$,
with vanishing entropy ($S \rightarrow 0$ and therefore a unique
state) warrants an elaboration.  We notice that there exist packings
above RCP all the way to $\phi_{\rm FCC} = 0.74048$, but these
packings have some degree of order.  These partially ordered packings
do not appear in our theory because we treat only disordered packings
characterized by the mesoscopic volume function which has been derived
under the isotropic assumption.  By doing so, we explicitly do not
consider crystals or partially crystalline packings in the ensemble.
This interprets the RCP in the context of the found third-law of
thermodynamics.  Our approach of neglecting the crystal state from the
ensemble has analogies in replica treatment of glasses \cite{zamponi}.

The mesoscopic entropy vanishes at RCP.
From this point the
entropy increases monotonically with $X$, being maximum for the
RLP limit. We note that the microscopic states contribute still to
the entropy of RCP giving rise to more states than predicted by
the present mesoscopic approach. This case is discussed in more
detail in \cite{jamming3}.

The equation of state $\phi(X)$ for different values of $Z$ can be
seen in the projection of Fig. \ref{entropy}. The volume fraction
diminishes with increasing compactivity according to the theoretical
picture of the phase diagram.  The curves $\phi(X)$ qualitatively
resemble the reversible branch of compaction curves in the experiments
of \cite{chicago2} for shaken granular materials and oscillatory
compression of grains \cite{sdr} suggesting a correspondence between
$X$ and shaking amplitude. The intention is that, different control
parameters in experiments could be related to a state variable, and
therefore might help experimentalists to describe results obtained
under different protocols.

For any value of $Z$, there is a common limit $\phi \to \phi_{\rm
  RCP}$ as $X \to 0$, indicating the constant volume fraction for all
the RCP states.  The singular nature of the frictionless $J$-point is
revealed as the volume fraction remains constant for any value of $X$,
explaining why this point is the confluence of the isocompactivity
lines, including RCP and RLP.  We conclude that at the frictionless
J-point the compactivity does not play a role, at least at the
mesoscopic level.

\subsection{Experimental realization of the phase diagram}

A reanalysis of the available experimental data tends to agree
with the above theoretical predictions.  However, it would be
desirable to perform more controlled experiments in light of the
present results.  As in other out of equilibrium systems, such as
glasses, the inherent path-dependency of jammed matter
materializes in the fact that different packing structures can be
realized with different preparation protocols
\cite{ciamarra,chicago2,swinney,sdr,bideau}
involving tapping, fluidized beds, settling particles at different
speeds, acoustic perturbations or pressure waves \cite{onoda}. Due to this reason, the value of $\phi_{\rm RCP}$ has not
been determined yet for monodisperse frictionless systems.
The more extensive evidence for a frictionless RCP appears from
simulations, which has been extensively performed for hard and
soft sphere systems. They find a common value of RCP for many
different preparation protocols \cite{makse,ohern,skoge}.
For frictional materials, experiments of Scott and Kilgour
\cite{scott}, and others display a nearly universal value of
volume fraction for RCP consistent with the theoretical
estimation.

Previous experiments \cite{onoda} and simulations \cite{zhang}
find that lower volume fractions can be achieved for smaller
settling speeds of the grains or slower compression (or quenching)
rates during packing preparation. Colloidal glasses find a similar
scenario \cite{parisi,kurchan}, with their glass transition
temperature dependence on the quench rate of cooling, although due
to different reasons.  This raises the question of whether the
jamming point is unique or determined by the preparation protocol
\cite{kurchan}.

The experiments performed by Onoda and Liniger with glass spheres
in liquid of varying density to adjust conditions of buoyancy in
the limit of vanishing gravity show that for larger values of
gravity, the volume fraction decreases.  This indicates that
settling speed of the particles can determine the final volume
fraction. Indeed, it was found numerically that the compression
rate during preparation of the packings is a systematic way to
obtain lower packing fractions, such that lower volume fractions
can be achieved with quasi-static compression rates during
preparation of the packings \cite{zhang}.

On the other hand, measuring the mechanical coordination
number in experiments seems to be an even more difficult task. In
the phase diagram, the RCP can be found at the frictionless point
with the highest possible coordination number $Z=6$.  Such a value
of $Z$ has been observed in the experiments of Bruji\'c {\it et
al.} on concentrated emulsions near the jamming transition
\cite{E21} and simulations of droplets.

The experiments of Bernal \cite{bernal}
simultaneously measured the coordination number and the
volume fraction.  Indeed, Mason, a postgraduate student of Bernal,
took on the task of shaking glass balls in a sack and freezing the
packing by pouring wax over the whole system. He would then
carefully take the packing apart, ball by ball, painstakingly
recording the positions of contacts for each of over $8000$
particles \cite{bernal}. Their result is a coordination number
$Z\approx 6$ and a volume fraction $\phi\approx 0.64$ which
corresponds to the prediction of the frictionless point in the
phase diagram. This may indicate that the particles used in the
experiment of Bernal have a low friction coefficient.

Another explanation, more plausible, is that the coordination
number measured by Bernal is not the mechanical one, but the
geometrical one. Indeed, we may argue that the only coordination
that can be measured in experiments of counting balls 'a la
Bernal' is the geometrical one, since, one is never sure if the
contacting balls were carrying a force or not.  Bernal could only
measure geometries, not forces.  This is in addition to the
uncertainty in the determination of the contacts with such a
rudimentary method as pouring wax and then counting one by one the
area of contact. The theory predicts that along the RCP-line, all
the RCP states have geometrical $z=6$, while $Z$ ranges from 4 to
6. Thus, it is quite plausible that the coordination number in
Bernal experiments is the geometrical one, $z=6$, in agreement
with the theory.

It is worth noting that since the labor-intensive method patented
half a century ago by Bernal, other groups have extracted data at
the level of the constituent particles using X-ray tomography
\cite{aste}.  Such experiments may give a clue to the relationship
between coordination number and volume fraction.  The experimental
data, to date, seems in good agreement with the present theory.
However, this method does not directly determine the contacts to
verify whether the particles are touching or just very close
together. Furthermore, the method cannot measure forces so the
distinction between geometrical and mechanical coordination is not
possible to achieve in this experiment.

An answer might be obtained using methods from biochemistry being
developed at the moment \cite{mukhopadhyay}. These methods promise to provide the high
resolution to determine the contact network with accuracy to
develop an experimental understanding of the problem.

An alternate way might be
generating the packings in the phase space through numerical
simulations, where both volume fraction and coordination number
could be easily determined.
Since $Z$ is directly determined by $\mu$, and the compactivity
determines what value of the volume fraction a packing has between
the limits of the phase diagram, the main question is how to
generate packings with different $\phi$ for a fixed $\mu$ to allow
the exploration of the phase diagram.

\section{Numerical tests}
\label{simulations}

In this section we perform two different numerical tests: on the
predictions of the theory and the assumptions of the theory. The
former are explained in Section \ref{results} while the latter are
elaborated in Section \ref{bounds}.  It is worth mentioning that while
the predictions can be tested with packings prepared numerically and
experimentally, the test of the assumptions of the theory is not so
trivial. This is because the theory is based on the existence of
quasiparticles which carry the information at a mesoscopic
scale. Thus, in principle we cannot use real packings, such as
computer generated packings or experimental ones, to measure the
quasiparticles. The information obtained from those packings already
contains ensemble averages through the Boltzmann distribution and
density of states. That is, it is in principle not possible to isolate
the behavior of quasiparticles from the real measures rendering
difficult to properly test some of the assumptions of the theory at
the more basic mesoscopic scale. 
In Section
\ref{bounds} we attempt to perform such a test, especially since we
can easily obtain the packings at the RLP line or the infinite
compactivity limit. These packings are fully random obtained as flat
averages in the ensemble without the corresponding Boltzmann factor
and may contain direct information on the mesoscopic fluctuations.

The existence of the theoretically inferred jammed states opens such
predictions to experimental and computational investigation.  We
numerically test the predictions of the phase diagram by preparing
monodisperse packings of Hertz-Mindlin \cite{landau,mindlin} spheres
with friction coefficient $\mu$ at the jamming transition using
methods previously developed \cite{makse,zhang,elastic,mgsj-pre}.

We test the theoretical predictions and show how to dynamically
generate all the packings in the phase space of configurations through
different preparation protocols.  Although diverse states are
predicted by the theory, they may not be easily accessible by
experimentation due to their low probability of occurrence.  For
instance, we will find that the packings close to the C-point (high
volume fraction, high friction, low coordination) are the most
difficult to obtain.

The advantage of our theoretical framework is that it systematically
classifies the different packings into a coherent picture of the phase
diagram.  In the following we use different preparation protocols to
generate all the phase space of jamming.  In particular we provide a
scheme to reproduce the RCP and RLP-lines amenable to experimental
tests.

We achieve different packing states by compressing a system from
an initial volume fraction $\phi_i$ with a compression rate
$\Gamma$ in a medium of viscosity (damping) $\eta$  where the
particles are dispersed. The system is defined by the friction
coefficient $\mu$ which sets $Z(\mu)$ according to Fig. \ref{z}.
While the simulations are not realistic (no gravity, boundaries,
or realistic protocol is employed), they provide a way to test the
main predictions of the theory.  The final state $(\phi,Z)$ is
achieved by the system for every $(\phi_i,\Gamma,\eta,\mu)$ at the
jamming transition of vanishing stress with a method explained
next.

It should be noted that other experimental protocols could be also
adapted to the exploration of the phase diagram, including (a) Gentle
tapping with servo mechanisms that adjust the system at a specified
pressure \cite{mgsj-pre}, (b) Gentle tapping with external oscillatory
perturbations, (c) Settling of grains under gravity in a variety of
liquids with viscosity $\eta$, (d) fluidized beds.


{\bf Relation with hard sphere simulations.---} The present
algorithm finds analogies with recent attempts to describe jamming
using ideas coming from the theory of mean-field spin glasses and
optimization problems \cite{parisi,kurchan}.  This is an
interesting situation since it has been shown that in the case of
hard spheres the system always crystallizes unless an infinite
quench is applied \cite{torquato}.

However, our soft sphere
simulations do not produce appreciable crystallization in the
packing.
This situation can be understood as following: In order to avoid the
(partial) crystallization in practice, a fast compression is needed. A
simple dimensionality analysis tells that the quenching rate $\Gamma
\propto G^{1/2}$, where $G$ is the shear modulus of the grains (see
Fig. \ref{G}). This analysis allows for comparison of our soft ball
simulations with the hard sphere simulations used in other studies
\cite{torquato}.
One could imagine that there exists a typical quench rate,
depending on the stiffness of the grains, as a division of the
ordered and disordered (jammed) phases. It is interesting to note
that at the hard-sphere limit, $G\to \infty$, all the packings are
in the ordered phase except when $\Gamma \rightarrow \infty$,
where $\phi\approx 0.64$. This explains the behavior of the hard
sphere packings found in previous works \cite{torquato}.  For the
soft ball case, the situation is more subtle since a certain range
of compression rate is acceptable for the acquirement of
disordered packings.  It is in this range that we perform
simulations. However, it should be noted that the above simple dimensionality analysis does not consider the dependence on dimensions and algorithms, which probably has important effects in the real situation.

According to this picture, in the theoretical limit of infinitely slow
quench rates, the system should in principle crystallize for any
$G$. However in practice, this limit is almost impossible to achieve experimentally. For sufficiently fast quench rates, crystallization can be avoided and a granular system is stuck in a random jammed configuration since its configurational relaxation time exceeds the laboratory timescale.
The contention is that, for a
well defined set of quenches, the system does not crystallize and the
study of the disordered phase ensues. Indeed, numerical and
experimental evidence points to the validity of this assertion. The
present theory neglects the existence of the ordered phase and assumes
an ensemble of the disordered states.  A more general theory will
account for the ordered states as well.  But this is a very difficult
problem in jamming: a full theory that accounts for RCP and
crystallization. A recent study has attempted to answer the question using Edwards' thermodynamic viewpoint of granular matter \cite{jin}.
From the simulation point of view, we can say that
our soft sphere approach does not produce crystallization while the
hard sphere approach does not produce randomness for any finite
compression rate.

\subsection{Molecular dynamics simulations of grains}
\label{md}

We prepare static packings of spherical grains interacting via elastic
forces and Coulomb friction (see \cite{makse,zhang} for more
details). The system size ranges from $N=1,024$ to $N=10,000$
particles. Two spherical grains in contact at positions $\vec{x}_1$
and $\vec{x}_2$ and with radius $R$ interact with a Hertz normal
repulsive force \cite{landau}

\begin{equation}
F_n = \frac{2}{3}~ k_n
R^{1/2}\delta ^{3/2},
\end{equation}
and an incremental Mindlin tangential force \cite{mindlin}

\begin{equation}
\Delta F_t= k_t (R
\delta)^{1/2} \Delta s,
\end{equation}
Here the normal overlap is $\delta= (1/2)[2 R - |\vec{x}_1 -
\vec{x}_2|]>0$. The normal force acts only in compression, $F_n =
0$ when $\delta<0$.  The variable $s$ is defined such that the
relative shear displacement between the two grain centers is $2s$.
The prefactors $k_n=4 G / (1-\nu)$ and $k_t = 8 G / (2-\nu)$ are
defined in terms of the shear modulus $G$ and the Poisson's ratio
$\nu$ of the material from which the grains are made. We use
$G=29$ GPa and $\nu = 0.2$ typical values for spherical glass
beads and we use $R=5\times 10^{-3}$ m and the density of the
particles, $\rho=2 \times 10^{3}$ kg/m$^3$. Viscous dissipative
forces are added at the global level affecting the total velocity
of each particle through a term $-\gamma \dot{\vec{x}}$ in the
equation of motion, where $\gamma$ is the damping coefficient
related to the viscosity of the medium $\eta = \gamma / (6 \pi
R)$.  Sliding friction is also considered:
\begin{equation}
F_t \le \mu F_n.
\end{equation}
That is, when $F_t$ exceeds the Coulomb threshold, $\mu F_n$, the grains slide and $F_t = \mu F_n$, where $\mu$ is the static friction coefficient between the spheres.  We measure the time in units of $t_0=R\sqrt{\rho/G}$, the compression rate in units of $\Gamma_0=5.9 t_0^{-1}$ and the viscosity in units of $\eta_0= 8.2 R^2\rho/t_0$. We choose the time step to be a fraction of the time unit $t_0$, which is the time that it takes for sound waves to propagate on one grain (see \cite{mgsj-pre} for more details). The dynamics follows integration of Newton's equation for translations and rotations.

\subsection{Preparation protocol: Packings at the jamming transition
  with the split algorithm}
\label{prep}


The critical volume fraction at the jamming transition, $\phi_{c}$
will be identified by the
``split'' algorithm as explained in \cite{zhang}, allowing one to
obtain packings at the critical density of jamming with arbitrary
precision. Preliminary results indicate that lower values of $\phi$
are obtained for slow compression \cite{zhang}. Thus we expect that by
varying $\Gamma$ or $\eta$ the phase space will be explored. Also, the
initial $\phi_i$ plays an important role in achieving packings close
to the RCP line.

The preparation protocol consists of first preparing a gas of
non-interacting particles at an initial volume fraction $\phi_i$ in a
periodically repeated cubic box.  The particles do not interact and
therefore the stress in the system is $\sigma=0$ and $Z=0$.


To generate a random configuration with friction at volume fraction
$\sim 0.64$ is very difficult.  Therefore to achieve any volume
fraction in this initial stage, we initially work with a frictionless
system.  We first generate a very dilute (unjammed gas of non
interacting particles) grain configuration randomly, usually with a
volume fraction $\phi_0 \approx 0.30 \sim 0.36$.  Then we apply an
extremely slow isotropic compression without friction on this dilute
configuration until the system reaches another unjammed configuration
at higher density, $\phi_i$ (see Fig. \ref{protocol}a). The reason to
add an extremely slow isotropic compression to the dilute
configuration is to avoid involving any kinetic energy to keep the
system as random as possible. After obtaining this unjammed state with
initial volume fraction $\phi_i$, we restore friction and a
compression is applied with a compression rate $\Gamma$ until a given
volume fraction $\phi_1$. Then the compression is stopped and the
system is allowed to relax to mechanical equilibrium by following
Newton's equations without further compression.

After the compression,
two things can occur (see Fig.
\ref{criticality}):

(a) The system jams: If the volume fraction $\phi_1$ is above
the jamming point, $\phi_1>\phi_c$, then the stress will
decrease and ultimately stabilize to a finite nonzero value,
meaning that the pressure of the system remains unchanged (usually
$\Delta \sigma < 10^{-3}$ Pa, red line, Fig \ref{criticality})
over a large period of time (usually $\sim 10^{7}$ MD steps). The
coordination number usually has a first initial decrease, but if
the system is jammed it will also stabilize at a constant value
above the isostatic minimal number (inset Fig. \ref{criticality}).

(b) The system is not jammed: If the volume fraction $\phi_2$
is below the jamming point, $\phi_2<\phi_c$, then the stress and
the coordination number will relax to zero.  This fact is
illustrated in Fig. \ref{criticality}.  If the packing has
$\phi_{1}>$ $\phi_{c}$, it stabilizes at a non-zero pressure above
the jamming transition, but the pressure decreases very quickly to
zero (the system is not jammed) if $\phi_{2}< \phi_{c}$, even
though $\phi_{1}$ and $\phi_{2}$ differ only by $2 \times
10^{-4}$.

We identify the exact volume fraction of the jamming transition
$\phi_c$, as follows \cite{zhang}:
A search procedure consisting of several cycles is applied such
that in each cycle we fix the lower and upper boundaries of
$\phi_{c}$, see Fig. \ref{chart}. The difference between the
boundaries gets smaller as the cycles proceed, meaning that
$\phi_{c}$ is fixed with higher and higher precision. We
start from the packing of high volume fraction $\phi_1>\phi_c$ and
generate a series of packings with step-decreasing volume
fractions until the first packing with zero pressure is observed,
which has a volume fraction $\phi_2=\phi_1-\Delta\phi$, where
$\Delta\phi$ is the difference between the jammmed packing
fraction $\phi_1$ and the unjammed packing fraction $\phi_2$.
Thus, $\phi_{c}$ is bounded between $\phi_1-\Delta\phi$ and
$\phi_1$. Then we test $\phi=\phi_1-\Delta\phi/2$. If
$\phi=\phi_1-\Delta\phi/2$ is stable, $\phi_{c}\ $is between
$\phi_1-\Delta\phi$ and $\phi_1-\Delta\phi/2$. If
$\phi=\phi_1-\Delta\phi/2$ is unstable, $\phi_{c}$ is between
$\phi_1-\Delta\phi/2$ and $\phi_1$. Therefore, in this cycle, we
reduce the region where $\phi_{c}$ possibly lies in from
$\Delta\phi$ to $\Delta\phi/2$. If we carry out this cycle for $n$
times, we improve the precision to
$\Delta\phi_n=\Delta\phi/2^{n}$. In our simulations cycling ceases
when $\Delta\phi_{n}$ gets below $2\times 10^{-4}$ and $n=12$. A
similar algorithm was employed in \cite{zhang} to study the
approach to the jamming transition by preparing packings at a
finite pressure. In the present work we are interested in jammed
packings at vanishing pressure, right at the jamming transition in
the isostatic plane defined for all friction coefficients.


It is important to determine whether the packings are jammed in the
sense that they are not only mechanically stable but also they are
stable under perturbations.
Our numerical protocols assure that the system is at least locally
jammed since each particle is in mechanical equilibrium
\cite{salvatore1}. To test if the system is collectively jammed is
more involved.  For frictionless systems, where tangential forces are
removed, we use the Hertz energy $U_{\rm hertz} = \frac{4}{15}~ k_n
R^{1/2}\delta ^{5/2}$ to test whether the Hessian of the jammed
configurations is positive \cite{ohern}. We find that the frictionless
configurations have positive Hessian indicating that they are
collectively jammed. However, this method is not useful when
considering frictional systems. The energy of deformation depends on
the path taken to deform the system and cannot be defined uniquely.
In this case, a numerical test of the stability of the packings
applies a small random velocity to each jammed particle.  We find that
the packings are stable to small perturbations consisting of external
forces of the order of 0.1 times the value of the average force,
indicating that the packings may be collectively jammed.


To test for more strict jammed conditions involves studying the
stability under boundary deformations.  We have tested that our
packings are stable under the most common strain deformations to
isotropic packings by performing a uniaxial compression test, a simple
shear and a pure shear test.

A simple shear test implies a strain deformation $\Delta \epsilon_{12}
= \Delta \epsilon_{21} \neq 0$, while the rest of the strain
components $\epsilon_{ij}$ remain constant.  A pure shear test is done
with $\Delta \epsilon_{11} = - \Delta \epsilon_{22}\neq 0$ and a
uniaxial compression test along the 1-direction is performed by
keeping the strain constant in $\Delta \epsilon_{22} = \Delta
\epsilon_{33} = 0$, and $\Delta \epsilon_{11}\neq 0$.  Here, the
strain $\epsilon_{ij}$, is determined from the imposed dimensions of
the unit cell. For example, $\epsilon_{11} = \Delta L /L_0$ where
$\Delta L$ is the infinitesimal change in the $11$ direction and $L_0$
is the size of the reference state.

In all cases the packings are stable under strain perturbations
indicative of the existence of a finite shear modulus (from the shear
and compression tests) and a finite bulk modulus (from the compression
test) of the jammed packings.  A full investigation of the elasticity
of the jamming phase diagram is left for future studies. It suffices
to state that the numerically found states can be considered to be
mechanically stable jammed states amenable to a statistical mechanical
description in the sense of the reversible branch of compaction in the
experiments of \cite{chicago2,sdr,bideau}.

\subsection{Results: phase diagram}
\label{results}

We repeat the above procedure 10 times with different random
initial configurations to get a better average of $\phi_c$. Figure
\ref{phaseb} shows the results for a system of $N=1024$.  All of
the packings shown in the present work as well as the code to
generate them are available at http://jamlab.org.  Each data point
corresponds to a single set of $(\phi_i, \Gamma, \eta, \mu)$ and
is averaged over these 10 realizations. We consider $0.40\le
\phi_i \le 0.63$, $10^{-7} \le \Gamma\le 10^{-3}$, $10^{-4} \le
\eta \le 10^{-3}$, and $0\le\mu \le \infty$. The error of $\phi_c$
obtained over the 10 realizations as shown in Fig. \ref{phaseb} is
usually $5\times 10^{-4}$, larger than the precision of the split
algorithm ($2 \times 10^{-4}$).  In general, all numerically
generated jammed states lie approximately within the predicted
bounds of the phase diagram.

\begin{figure}
\centering{
\vbox {
 \resizebox{9cm}{!}{\includegraphics{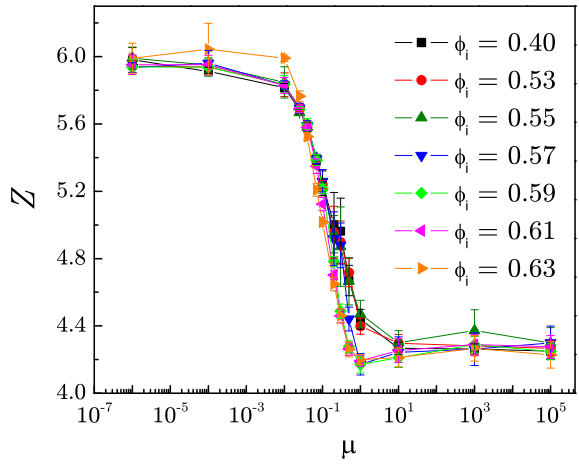}}
}
}
\caption{Mechanical coordination number versus friction $\mu$ obtained
  in our numerical simulations explained in Section \ref{simulations}
  for different preparation protocols characterized by the initial
  volume fractions $\phi_i$ indicated in the figure. The symbols and
  parameters used in these simulations are the same as in the plot of
  Fig.  \ref{phaseb}.}
\label{z}
\end{figure}
\begin{figure}
  \centering {
      \resizebox{6cm}{!}{\includegraphics{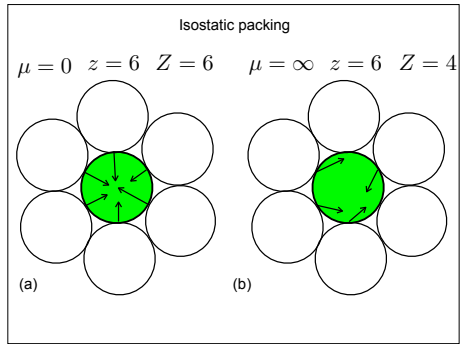}}
    }
\caption{ (a) Consider a frictionless packing at the isostatic
    limit with $z=6$.  In this case the isostatic condition implies
    also $Z=6$ mechanical forces from the surrounding particles. (b)
    If we now switch on the tangential forces using the same packing
    as in (a) by setting $\mu\to\infty$, the particle requires only
    $Z=4$ contacts to be rigid. Such a solution is guaranteed by the
    isostatic condition for $\mu\to\infty$. Thus, the particle still
    have $z=6$ geometrical neighbors but only $Z=4$ mechanical ones. }
\label{possible}
\end{figure}

The plot in Fig. \ref{phaseb}a explores the dependence of the packings
$(\phi,Z)$ on the initial state $\phi_i$. (In the plot, $\phi$ refers
to the $\phi_c$ obtained in a split algorithm).  In Fig. \ref{phaseb}
we plot our results for a fixed quench rate $\Gamma=10^{-7}$ and
damping coefficient $\eta=10^{-3}$ (except for the last orange curve
on the right with $\eta=10^{-4}$) and for different initial states
ranging from left to right (see Fig. \ref{phaseb} for details)
Each color here corresponds to a particular $\phi_i$ (0.40 at the left
to 0.63 at the right), while each data point along a curve corresponds
to a system prepared at a different friction from $\mu=0$ at $Z\approx
6$ to $\mu\to \infty$ at $Z\approx 4$.  For instance, the states
joined by the dashed lines in Fig.  \ref{phaseb}a are for systems with
the same $\mu $. While these lines are approximately horizontal,
indicating an approximate independence of $Z$ on the preparation
protocol $\phi_i$, we observe deviations specially around the RLP line
for intermediate values of $\mu$.  Despite this, the $Z(\mu)$
dependence is approximately valid as seen in Fig. \ref{z}, which is
plotted using the results of Fig. \ref{phaseb}a.

We find that the packings prepared from the larger initial densities
$\phi_i$ closely reproduce the RCP line of zero compactivity at
$\phi_{\rm RCP}$, confirming the prediction that RCP extend along the
vertical line from the J-point to the C-point.

We find that the packings along the RCP line have equal geometrical
coordination number $z\approx 6$ but differ in their mechanical one
from $Z=6$ to $Z\approx 4$, in agreement with theory (see Section
\ref{bounds}).  These states are then identified with the ground state
of jammed matter for a given $\mu$ as depicted in the volume landscape
picture of Fig. \ref{landscape}.


In the limit of small initial densities (see curves for $\phi_i=0.40,
0.53$ and $0.55$ in Fig.  \ref{phaseb}a) we reproduce approximately the
predictions of the RLP-line.  These packings follow the theoretical
prediction for infinite compactivity except for deviations (less than
5\%) in the coordination number for packings close to the lower value
of $Z=4$, where the numerical curve seems to flatten out deviating
from the theory.  Numerically, we find the lowest volume fraction at
$\phi_{\rm RLP}^{\rm min}=0.539 \pm 0.003$, close to the theoretical
prediction, $\phi_{\rm RLP}^{\rm min}=0.536$. We notice that the
theoretical lines in Fig. \ref{phaseb} are obtained for
$h_z=e^{-100}$, thus the minimum RLP deviates slightly from the
theoretical prediction in the limit $h_z\to 0$ as discussed in Section
\ref{phase-diagram}.


Packings with the lowest $Z$ correspond to infinitely rough spheres.
The deviation of coordination number between theory and simulation
(specially at low volume fraction) could be from the system not
achieving an isostatic state at infinite friction.  We were not able
to generate packings with exactly $Z=4$ but our algorithm generates
packings very close to this isostatic packings at $Z\approx 4.2$ for
$\mu=10^4$. In general, we find that while there are many states along
the RLP line, the barriers of these states decrease as the
coordination number decreases towards $Z=4$, i.e. when the friction
increases.  Thus, the states at the lower left part of the phase
diagram are very difficult to equilibrate.

Besides the states delimiting the phase space, we generate other
packings with intermediate values of $\phi_i = 0.57, 0.59, 0.61$ as
shown in Fig. \ref{phaseb}a.  Interestingly, we find that these states
(all obtained for fixed $\Gamma=10^{-7}$ and $\eta=10^{-3}$) closely
follow the predicted lines of isocompactivity as indicated in the
figure.  We find that the simulations from $\phi_i = 0.57, 0.59, 0.61$
correspond to compactivities $X=1.62, 1.38, $ and 1.16, respectively
(measured in units of $10^{-3}V_g$).

The constant $h_z$
weakly affects the finite compactivity lines.  We find that
$h_z=e^{-100}$ provides the best fit to the data for finite
isocompactivity lines in Fig. \ref{phaseb}a.  Thus, with
reasonable approximation and for this particular protocol, we
identify the density of the initial state, $\phi_i$ with the
compactivity of the packing, providing a way to prepare a packing
with a desired compactivity.  We note, however, that other values
of $h_z$ produce approximately the same phase diagram boundaries
(i.e., the RLP and RCP lines, as long as $h_z\ll 1$) but do not
fit the isocompactivity lines within as accurately. Thus, the
identification of the compactivity with $\phi_i$ remains dependent
on this particular value of $h_z$ used to fit the theory.

%

We also test the dependence on the state of the packings with the
compression rate $\Gamma$ and viscosity $\eta$. Both parameters should
have similar effects since they slow down the dynamics of the grains.
Figures \ref{protocol}b and \ref{protocol}c represent typical
preparations.  In general, we reproduce the RLP line for slow quenches
or for large viscosities as seen in Fig. \ref{phaseb}b (for $\phi_i =
0.40$ in black).
In this case, the grains are allowed to slide and develop large
transverse displacements and Mindlin forces with the concomitant low
$\phi$ and large compactivity.  Therefore we find a predominance of
the Mindlin forces over the normal Herztian forces as a characteristic
of the lower volume fractions of RLP, having a pronounced path
dependent structure with large compactivity in the packing.  When the
compression rate is increased or the damping is reduced we find
packings with higher volume fractions as indicated in
Fig. \ref{phaseb}b.  Indeed, analogous protocol to generate these
packings would be to change the damping coefficient by immersing the
particles in liquids of different viscosities.



It is also interesting to investigate the packings generated by fixed
friction and varying quenched rate, since this situation corresponds
to a given system.  For a given $\mu$ we find approximately a common
value of $Z(\mu)$ independent of $\Gamma$ and $\phi_i$, as discussed
in Fig.  \ref{z}, indicating that the curve $Z(\mu)$ is roughly
independent of the protocol. Thus, a fixed $\mu$ corresponds to a
horizontal line in the phase diagram as indicated in Fig \ref{phaseb}a
by the dashed lines.




Figure \ref{protocol} shows a summary of the numerical protocols we
use to generate the packings in the phase diagram. The ``easiest''
packings to generate are along the RLP line (Fig.  \ref{protocol}b, c)
while the most difficult are deep in the phase diagram with low
coordination number (large friction) and high volume fraction, that is
near the C-point (Fig. \ref{protocol}a).  In general, we always start
with a dilute unjammed (non-interacting gas) sample without friction
and compressed slowly until another dilute sample not jammed at a
volume fraction $\phi_i$. If $\phi_i$ is around or smaller than $0.55$
then the RLP line is obtained. If we decrease by 3 orders of
magnitudes the compression rate using this initial density, we move
inside the phase diagram, though not substantially, as indicated
schematically in the figure.

Therefore, in order to access the inner states around the C-point at
high friction, we took another route. We prepared a frictionless
dilute unjammed sample at a higher $\phi_i$, which can be as close as
possible to $\sim 0.63$ since frictionless spheres only jam at
RCP. Then we switched on friction and got the packings inside the
phase diagram as indicated in Fig. \ref{protocol}a. We realize that
switching on and off friction is impossible in experiments. However, a
possible way to experimentally realize the path to the C-point is by
resetting the path dependency to zero along the preparation protocol:
this path could be reproduced by stabilizing a frictional packing with
fast compressions and resetting the interparticle forces by gently
tapping. This will simulated the lost of path dependence that we
simulate by the initial compression without friction and may allow to
reach the packings around the C-point.


\subsection{ Compactimeter for granular matter and the ABC experiment}
\label{compactimeter}

At this point we do not have a theoretical explanation for why
packings with the same initial unjammed state $\phi_i$ end up having
the same compactivity when jammed after compression.  It is important
to note that this finding is based on the particular value we use for
$h_{z}$, such that a change in $h_{z}$ gives rise to different lines
of isocompactivity which may not fit the simulation results of
Fig. \ref{phaseb}a so closely.  Nevertheless, the close fit between
theory and simulation deserves an explanation.

We may conjecture that $\phi_i$ determines a type of disorder quenched
in the initial configuration that leads to systems with the same
compactivity but different volume fractions and coordination numbers,
evidenced by our results.  We can use this empirical result to control
the compactivity of the packing, at least for this particular
protocol, defining a ``granular compactimeter''. Laboratory
measurements of compactivity usually involve indirect measures through
Fluctuation-Dissipation relations between fluctuations and response
functions \cite{chicago2,swinney,mk}. However, there is no simple
thermometer (or "compactimeter") to measure this observable directly
in a packing.  Furthermore, there is no simple way to prepare a
packing with a desire compactivity. Our results can be used to, at a
minimum, define packings with equal compactivity. The empirical
identification of $X$ with $\phi_i$ promotes the possibility of
controlling $X$ within this particular protocol, an important step in
any thermodynamical analysis.

The advance of a granular thermodynamics crucially depends on the
invention of a thermometer that can easily measure the
compactivity, as in Fig. \ref{thermometer}.
A zero-th law of thermodynamics presents serious challenges in
granular materials since there is no straight forward way to
define a compactivity bath or reservoir.

The present theory predicts the dependence of the volume fraction,
$\phi(X)$, on the compactivity, allowing for the development of a
compactimeter: Such a device consists of a known granular medium,
having a given $\mu$ and, therefore, a given coordination number.  The
compactimeter is inserted in a granular system and both, system and
compactimeter, are gently shaken to allow for equilibration at a
common compactivity (the compactimeter has a flexible membrane that
allows volume transfer between both systems in contact).
The volume of the grain in the compactimeter will adjust accordingly
and the compactivity can be read directly from the scale attached to
it through the equation of state, $X(\phi)$, provided by the theory.

The foundation of the above compactimeter relies on the validity
of a zero-th law for granular thermodynamics that determines the
equilibration at a given compactivity of a system:  Two systems in
contact should mechanically equilibrate at the same compactivity. A recent experiment \cite{lechenault} found a materials-independent relationship between the average volume fraction and its fluctuations in two equilibrated granular subsystems, which gives support to the zero-th law. Further test of the zero-th law is difficult to facilitate without
theoretical guidance. Thus, it will be useful to perform the
following test, either numerically or experimentally to test not
only the idea of equilibration but also the possibility to
describe granular matter under the V-ensemble. Such a test has
been labeled the ABC experiment as proposed by Edwards \cite{E8}, which is now possible to perform using the phase diagram.

Two packings are numerically prepared at points A and B in the
Fig. \ref{abc} for different $\mu$, $\mu_{\rm A}<\mu_{\rm B}$. The
systems are equilibrated at different compactivities $X_{\rm A} <
X_{\rm B}$ by using different $\phi_i$. They are then put into contact
through a flexible membrane (or by simply putting the particles at the
surface in contact) and allowed to mechanically equilibrate by gently
shaking the system. Numerically, this can be simulated with the same
compression methods as explained in Section \ref{simulations}. The
total volume is kept fixed and the volumes of the subsystems should
change accordingly, $V = V_A + V_B$.  If the AB system equilibrates at
the same compactivity, then it will follow the trajectory depicted in
Fig. \ref{abc} towards the state C of equilibration along the
isocompactivity line. There is no need to measure the compactivity of
the final packing. By just measuring the final volumes $V_A$ and $V_B$
(or their volume fractions) will suffice. This will provide an
important test for the V-ensemble and compactivity to describe jammed
matter, as well as the validity of the phase diagram.



Another intriguing possibility would be to mix a frictionless A
packing at the J-point with an infinite friction B packing at the
L-point, which can be achieved numerically.  This AB system would
provide the maximum difference between the isolated packings, allowing
study of the zero-th law with more accuracy.  According to theory, the
frictionless packing is independent of $X$. Therefore, it should
remain at the J-point. The $\mu\to\infty$ packing should, in
principle, equilibrate at any $X$ or volume fraction along the
G-line. Most probably, it would stay put at the L-point since the A
and B packings are joined by the isocompactivity $X\to \infty$ line
already. In any case, the resulting AB packing should fall inside the
phase diagram if the theory is correct. If it does not equilibrate
(falling outside the phase diagram, we note that the two packings are
at the borderline), it could indicate the breakdown of the approach.
Two conclusions could be reached from the failure of such a
test. Either the V-ensemble is wrong or the approximations of the
V-ensemble theory are not correct. In the latter, more sophisticated
theories should be developed.  If the former case occurs, then a
different ensemble will have to be considered beyond isostaticity.

\subsection{Microscopic states versus mesoscopic states}
\label{micro_states}

The notion that RCP arises for $X=0$, implying that there are no
fluctuations at RCP, deserves some discussion. Furthermore, it is
important to discuss the implications of the related coincidence of
RCP and RLP at the frictionless point. That is, the fact that for
frictionless particles there is a unique state of jamming and
therefore the compactivity is irrelevant when $\mu=0$.

First of all, we need to emphasize that these results should be
interpreted at the mesoscopic level. Thus, there is one mesoscopic
state at RCP and only the mesoscopic entropy vanishes at this point.
These results provides an unambiguous interpretation of RCP only at
the meso level.  As discussed above, for a single mesoscopic state,
there are many microscopic states average out in the calculation of
the average free volume function $w(z)$ which is used in the partition
function Eq. (\ref{pf1}) (see also Jamming I
\cite{jamming1}). Therefore, we expect that these microscopic states
will contribute to the entropy of RCP (see Jamming III \cite{jamming3} for a
discussion) as well as giving rise to other states perhaps with
different volume fraction.  Below we elaborate on the existence of
microscopic and mesoscopic packing states.  This point is better
discussed in terms of the volume landscape of Fig. \ref{landscape}:
all the jammed states are degenerated around the mesoscopic ground
state with the coordination number $z=6$. These states have slightly
different volume fractions, leading to microscopic fluctuations which
are coarse-grained in the mesoscopic theory.
The existence of microscopic
states neglected at the mesoscopic level
raises an interesting analogy with recent work of Zamponi and
Parisi \cite{parisi} where the jammed states are considered as
infinite pressure glassy states obtained after a fast compression from
a liquid state at finite temperature.  They propose that a range of
different volume fractions of jamming can be achieved according to the
initial state from where the quench is started, with the maximum
possible jammed density obtained when the initial density is the
Kauzmann point of the ideal glass.  In their representation (see
Fig. 4 in \cite{parisi}) a range of initial
densities in the liquid phase $\varphi \in [\varphi_d,\varphi_K]$,
where $ \varphi_d$ is the density where many metastable states first
appear in the liquid phase as suggested by mean field model picture of
the glass transition, and $ \varphi_K$ is the Kauzmann density of the
ideal glass transition, gives rise to a final density of jamming,
$\varphi_j(\varphi)$, obtained after a fast quench from $\varphi$.
Two limiting densities of jamming are predicted: the minimum at
$\varphi_j(\varphi_d) = \varphi_{th}$ and the maximum volume fraction
obtained after a quench from the Kauzmann point, $\varphi_j(\varphi_K)
= \varphi_{0}$.

This situation could be interpreted in term of neglecting the microscopic states in our theory.
It is also a question whether the states at infinite pressure obtained using fast quenches from a liquid state ``a la Zamponi'' are
equivalent to the jammed state obtained in our approach for soft spheres. Our volume ensemble approach might differ from the energy/force ensemble used in the work of
Zamponi-Parisi. Further investigations are required to reveal the possible differences. 

\subsection{Geometrical versus mechanical coordination number: measuring the behavior of quasiparticles}
\label{bounds}

In Section \ref{results} we have tested the predictions of the
theory. In this Section we attempt to test some of the assumptions of
the theory, specially those related to the quasiparticles of $w(z)$
and the bounds in $z$ and its distribution.  A quasiparticle is a
mathematical entity that is, in principle, impossible to isolate from
the structure of a real packing prepared either with MD computer
simulations or experiments, rendering the measurement of their
properties from real packings difficult. This is because real packings
already contain the ensemble average, and we can not dissociate the
ensemble, with the Boltzmann distribution of states, compactivity and
density of states, from the isolated statistics of a quasiparticle.

Quasiparticles can be tested by preparing random packings with
prescribed geometrical coordination number.  However, below, we argue
that we may obtain approximate properties from real packings in the
limit $X\to\infty$ and $h_z\to 0$ based on the following observations:
First, when $X\to \infty$ the Boltzmann factor is one, and the average
in the partition function is flat over the configurations up to the
density of states. Furthermore, the density of states is found to be a
very fast decaying function of $z$.  These two results implies that
along the RLP-line of real packings (either numerical or experimental)
we can obtain fully random packings which could reveal the properties
of the quasiparticles. Under this approximation we can then test some
of the approximations of the theory.  However, the conclusions from
this section remain approximate.

We have proposed that the bounds of the geometrical coordination
number are $Z\le z \le6$ and below we test these bounds.
We find numerically that the quasiparticles size is of the order of
two particle diameters.  Once $z$ is averaged over this range, we find
that $z$ ranges approximately between $(Z, 6)$.
In fact, we find even more narrow distribution than theoretically
expected. This is due to the fact that we define $z$ without the
microscopic fluctuations but including the mesoscopic
fluctuations. Once we average over its neighbors, we not only remove
the microscopic fluctuations but also, partially, the mesoscopic ones.
In the limiting case of no mesoscopic fluctuations, when $X = 0$ along
the RCP line, we find a very narrow distribution at $z \approx 6$ for
any value of $Z$ and $\mu$, after averaging $z$ over a mesoscopic
region of two particle diameters. The distribution is even narrower
when $z$ is coarse-grained over a region of four particle diameters.

The X-ray tomography experiments of Aste \cite{aste} (Fig. 6a) reveal
the trend predicted by Eq. (\ref{hamil}) between the inverse of the
Voronoi volume and the number of neighbors of a set of Voronoi cells
with similar volumes. However, it is evident from the figure that such
number of neighbors is spanning a range of values between $\sim 4$ and
$\sim 10$. The reason why the Aste's group \cite{aste} found a wider
range of $z$ is possibly due to the fact they did not consider a range
of coarse-graining.


A question arise as how to identify the geometrical from the
mechanical coordination number, even numerically. A detailed discussion of this topic appears in Ref. \cite{song_long}. Below we summarize the results. Strictly speaking,
this distinction is not possible to materialize in real packings.
Again, we need to generate random packings from a geometrical point of
view, without resorting to forces, and then study their distribution
in real packings where forces are taken into account.  Nevertheless,
we follow the following approximate scheme to try to obtain
information of the quasiparticles from real packings.

Even in numerical simulations, there is always a round off error
associated to measurements.  Thus, two particles that may not be in
contact (giving rise to a zero force) may be close enough to be
considered as contributing to the geometrical coordination.  Indeed,
it is known that $g(r)$ has a singularity \cite{donev}, $g(r) \sim
(r-.5)^{-0.5}$, implying that there are many particles almost
touching.
Following this consideration, we introduce a modified radial
distribution function (RDF) $g_z(r)$ in order to approximately
identify $z$ and $Z$ from real packings:

\begin{equation}\label{dgz}
  g_z(r)=\frac{1}{N}\frac{R^2}{r^2}\sum^N_i\sum^N_{j\neq
    i}\Theta\Big(\frac{r_{ij}}{r-R}-1\Big)\Theta\Big(\frac{r+R}{r_{ij}}-1\Big), \ \ r > R
\end{equation}
where $R$ is the radius of particle, $N$ is the number of particles,
$r_{ij}$ is the distance of two particle's centers,
$r_{ij}=|\vec{r}_i-\vec{r}_j|$, and $\Theta$ is the Heaviside step
function. The RDF describes the average value of the number of grains
in contact with a virtual particle of radius $r$, and the factor of
$R^2/r^2$ is the ratio of a real sphere's area and the virtual one's.

$g_z(r)$ measures the number of balls with their volume intersecting
the surface of a sphere of radius $r$ measured from the center of a
given ball.  When $r=R$ in (\ref{dgz}) we obtain the mechanical
coordination number while the geometrical one is obtained for a small
value $\Delta r = \frac{r-R}{2R} \ne 0$ for which we distinctly find a
signature from computer simulations, unambiguously defining it at
$\Delta r = 0.04$.

Figures \ref{gzr_RCP} and \ref{gzr_RLP} plot the $g_z(\Delta r)$ of
packings with various friction coefficient $\mu$ on the isostatic
plane along the RCP and RLP lines respectively.  Following the
definition of Eq. (\ref{dgz}), $g_z$, with $\Delta r = 0$, should be
directly equal to the mechanical coordination number, $Z$, and should
range from $4$ to $6$ along both RCP and RLP (if $h_z\ll1$) lines
which is confirmed by our numerical simulations in Figs. \ref{gzr_RCP}
and \ref{gzr_RLP}, respectively.

Furthermore, as shown in the figures, we find that $g_z(\Delta r)$
along the RCP line, increases slightly as $\Delta r$ increases, and
finally reach the same value of $g_z$ at $\Delta r = 0.04$ as shown in
Fig. \ref{gzr_RCP}. We identify the geometrical coordination number as
$z=g_z(0.04)$ under the accuracy of the simulations. Therefore, we
conclude that all RCP states have approximately the same geometrical
coordination number, $z\approx 6$, in agreement with the theory.  In
terms of the volume landscape of Fig. \ref{landscape}, the states
along the RCP line are ground states with different friction at
$X=0$. Thus they should all have $z=6$.  We notice that $g_z(\Delta
r=0.04)=6.65 > 6$, which may result, firstly, from the approximate
nature of our measurements of $z$ from real packings since we are not
measuring the quasiparticles directly, and, secondly, from the
increasing of the coordination number slightly away from the critical
point as the local volume fraction $\phi$ increases slightly as
$Z-Z_c\sim (\phi-\phi_c)^\beta$. We note that $Z$ is also slightly
increased from the isostatic point.

Along the RLP line, Fig. \ref{gzr_RLP}, we find that the geometrical
coordination number as extracted from $g_z(0.04)$ is very close to the
mechanical one.  Since the RLP line is at $X\to \infty$ and $h_z\ll
1$, all the states along RLP have $z\approx Z$ as we move along the
line varying the friction coefficient.  Thus, the numerical results
confirm the theory, and further confirm that the value of $h_z$ is
very small.

Next, we study the coarse-grained coordination number $\langle
z\rangle_l$ as defined as,

\begin{equation}
  \langle
  z\rangle_l=\frac{1}{N_l}\frac{R^2}{r^2}\sum^{N_l}_i\sum^{N_l}_{j\neq
    i}\Theta\Big(\frac{r_{ij}}{r-R}-1\Big)\Theta\Big(\frac{r+R}{r_{ij}}-1\Big), \ \ r > R
\end{equation}
where $N_l$ is the number of the particles inside a coarse-grained
spherical range with a radius of $l$. Figures \ref{cgz_RCP} and
\ref{cgz_RLP} plot the PDF of $\langle z\rangle_l$ for all the
packings along the RCP and RLP lines, respectively. The distributions
show a narrow shape, and the average value gives the value of
$g_z(r)$, i.e.,

\begin{equation}
\overline{\langle z\rangle_l}=g_z(r).
\end{equation}
We find that all the $P(\langle z\rangle_l)$ along the RCP line
coincide at $\Delta r = 0.04$ as shown in Fig. \ref{cgz_RCP}b and
\ref{cgz_RCP}c, demonstrating that all the states have approximately
the same average value of $z$ (as discussed above), as well as the
same distribution. The distribution gets narrower as the
coarse-graining parameter $l$ increase from $l=2$ to $l=4$ as shown in
Fig. \ref{cgz_RCP}c.  On the other hand $Z$ changes as the friction
changes following the isostatic condition, as seen in
Fig. \ref{cgz_RCP}a.  This in in good agreement with the theory.

Along the RLP line, the situation is analogous.  As discussed above,
$z\approx Z$. This is clearly demonstrated in Fig. \ref{cgz_RLP}b
which should be compared with the analogous Fig. \ref{cgz_RCP}b for
the RCP line. Figure \ref{cgz_RLP}a also shows how $Z$ changes with
friction in the same way as in Fig. \ref{cgz_RCP}a.  More importantly,
we find that the bounds of $\langle z\rangle_l$ are well approximated
between the bounds proposed by the theory $(Z,6)$.  The slight shift
towards higher coordinations observed in Fig. \ref{cgz_RLP}b is due to
the approximate nature of our measurements from real packings and the
small increment in volume fraction from the critical point of jamming,
as discussed above.  Overall, we conclude that along the RCP line, all
states have $z \approx 6$, with $Z$ changing from 6 to 4 as friction
is increased.  On the other hand, along the RLP line, we find that $z
\approx Z$.  These two results confirm the theory very well.  Figure
\ref{z-summary} summarizes all these finding indicating the ranges of
$z$ and $Z$ along the iso-$X$ lines of RCP and RLP, the iso-$z$
vertical lines, and the iso-$Z$ horizontal lines in the phase diagram.

\section{Outlook}

We have presented several results in the disordered sphere packing
problem and here will discuss their significance.  We start with
the results that we believe are more robust and may survive higher
scrutiny, and follow with the results that require further study
plus a discussion on how to improve the theory.

The final result of the phase diagram, the characterization of RCP
and RLP using the compactivity, the role of friction in the
determination of the limiting packings are interesting results,
almost independent of the theory. The distinction between the
geometrical and mechanical coordination number is also important,
allowing for the implementation of averages in the partition
function.  On the mathematical side, the microscopic volume
function and the formula obtained for the Voronoi cell in Jamming
I \cite{jamming1} are important and exact results which are the
foundation of all preceding work presented herein.

While the predictions Eqs. (\ref{RCP_state}) and (\ref{rlp_min}) of RCP and
RLP approximate very well the known values, they are obviously not
exact expressions, since they are based on several approximations of
the statistical theory of volumes. Thus, the implementation of the
mesoscopic volume function and the probability distribution of volumes
on which these results are based are, perhaps, the points where
improvements may be necessary. While the $\frac {2\sqrt{3}}{z}$
formula reproduces the average volume fraction surprisingly well, it
is based on approximations to calculate the probability distribution
of volumes, $P_{>}(c)$, as explained in Jamming I \cite{jamming1}.
More work is needed to obtain better approximations to such a
distribution to capture not only the mean value of the Voronoi volume
but higher moments as well. A full discussion is performed in Jamming
IV \cite{jamming4}. The present theory shows the way to improve upon
Jamming I and obtain an exact formulation of $P_{>}(c)$, at least to a
prescribed value of the coordination shell.  Such a formulation is
presently being developed and is based on a systematic study of
quasiparticles with fixed geometrical coordination number; it produces
good fits of $P({\cal W}_i^s)$ for the case of 2d, while 3d
calculations are under way.

Other approximations that require additional attention are the
considerations leading to the density of states $g(z)$.  While the
main results of RCP and RLP are nearly independent of the
particular form of $g(z)$, other quantities such as the entropy
and the equations of state for finite nonzero compactivity are
more sensitive to it. By a direct measurement of the entropy,
$g(z)$ can be calculated and compared to the simple exponential
form used in the present study.

Furthermore, we have explicitly removed crystal states from our
calculations by considering the upper bound of $z=6$. A more general
theory is needed then to investigate the transition from the RCP to
the FCC as crystals starts to form in the system. Such a theory will
include not only disordered states but also partially crystallized
states and would give $X=0$ at FCC following the full Edwards
partition function, Eq. (\ref{Q}). We investigate this idea in \cite{jin}, where we find that RCP can be interpreted as the ``freezing point" in a first-order phase transition between ordered and disordered packing phases.


The distribution of coordination number, assumed to be a delta
function, is another approximation to be improved upon.  Once again,
this may not affect the predicted RCP and RLP but deeper studies
require also a distribution of $z$.  A more general ensemble is being
considered that predicts the distribution of $z$ in good agreement
with the existing data (see Jamming IV \cite{jamming4}).

Finally, the extension to study the microstates requires more
investigation. The identification of the elementary units as
quasiparticles of fixed coordination number is interesting and
could eventually lead to more precise solutions of RCP and RLP.

\section{Conclusions}

In conclusion, using Edwards statistical mechanics we have elucidated
some aspects of RLP and RCP in the disordered spherical packing
problem.
The numerical results suggest a way to experimentally test the
existence of the predicted packings. By allowing the grains to settle
in liquids of varying density, the speed of the particles can be
varied and a systematic exploration of the jamming phase diagram can
be achieved.

\begin{figure}
\centering {
\vbox {
 \resizebox{9cm}{!}{\includegraphics{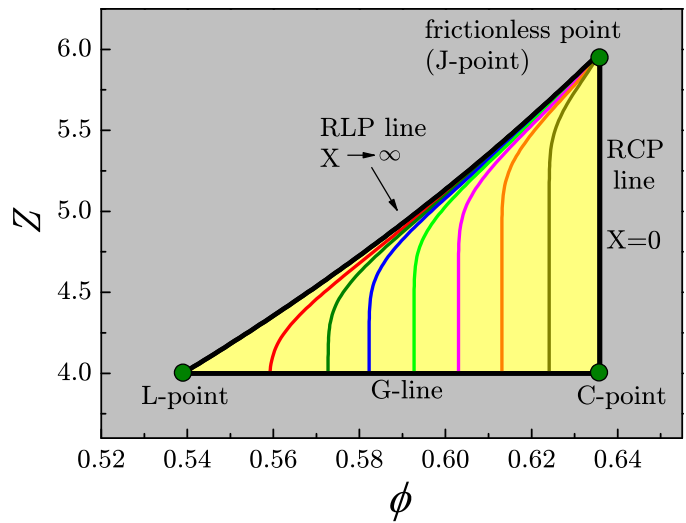}}
}
}
\caption{{\bf Phase diagram of jammed matter: Theory}.  Theoretical
  prediction of the statistical theory. All disordered packings lie
  within the yellow triangle demarcated by the RCP line, RLP line and
  G line. Lines of finite isocompactivity are in color. The grey area
  is the forbidden zone where no jammed packings can exist.}
\label{phasea}
\end{figure}

\begin{figure*}
\centering {
\vbox {
 \resizebox{12cm}{!}{\includegraphics{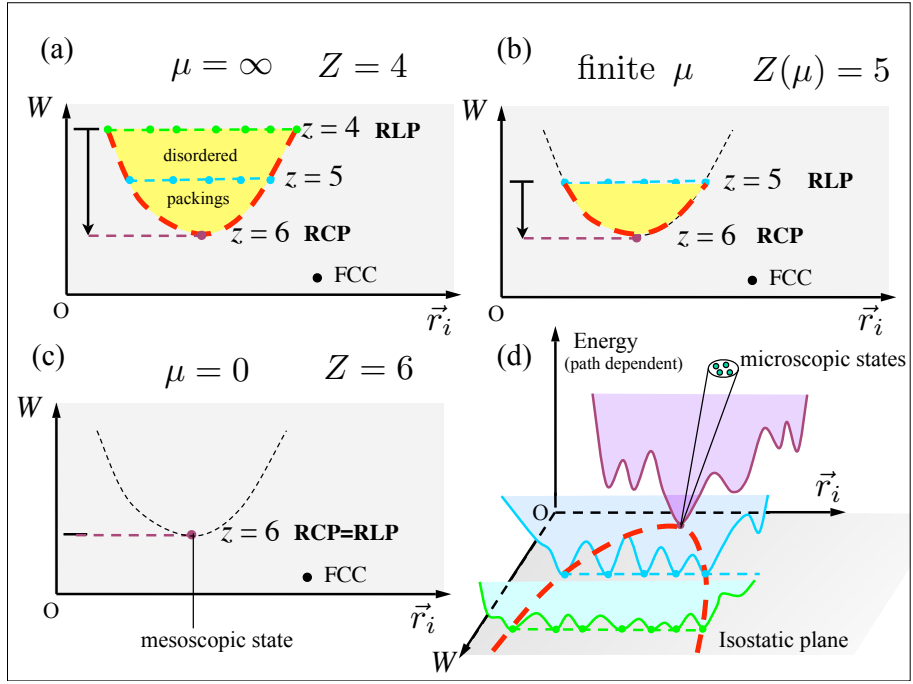}}
}
}
\caption{ {\bf Schematic representation of the volume landscape of
    jammed matter $(\vec{r}_i, W)$.}  The multidimensional coordinate
  $\vec{r}_i$ represents the degrees of freedom: the particle
  positions.  Each dot represents a discrete mesoscopic jammed states
  at different $z$. It is important to note that for each meso state
  there are microscopic states with the same $z$.  All disordered
  packings are in the yellow region (of an schematic shape) of the
  phase space which corresponds to the isostatic plane of hard spheres
  at the jamming transition where our calculations are
  performed. Other ordered packings have lower volume, such as the
  FCC.  (a) We represent the case of $\mu=\infty$. The states
  represent those along the G-line in Fig. \ref{phasea} as the
  compactivity varies from $X=0$ (ground state) to $X\to \infty$ at
  the RLP.  The horizontal lines indicate packings at constant
  volume. The ground state of jammed matter for this friction
  coefficient has $z=6$ and the highest volume states are found for
  $z=4$. The arrow indicates the limits of integration in the
  partition function for this particular friction. (b) For another
  finite $\mu$, the space is delimited by above by a line of constant
  $z=Z(\mu)$. (c) For $\mu = 0$ only the ground state is available,
  giving rise to a single state.  (d) The volume states states in
  (a,b,c) are separated by energy barriers represented by the third
  coordinate in the phase space.  The energy barriers are path
  dependent due to friction between the particles.  Nevertheless, the
  jammed configurations are well-defined in the isostatic plane, and
  the energy barriers represent the work done to go from one
  configuration to another.  Only at the frictionless point, the
  energy barriers are path independent. }
\label{landscape}
\end{figure*}

\begin{figure}
\centering {
\resizebox{9cm}{!}{\includegraphics{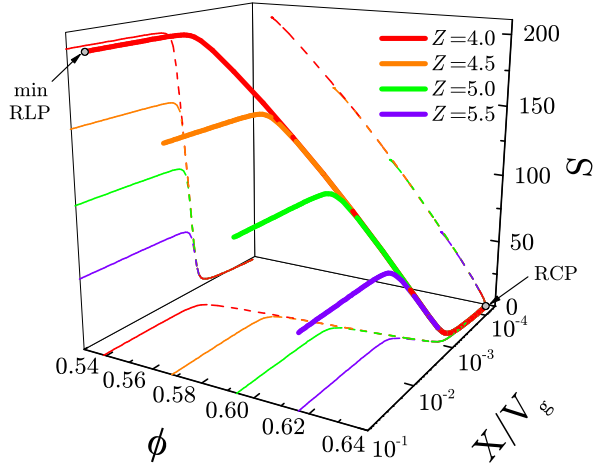}}
}
\caption{{\bf Predictions of the equation of state of jammed matter in
    the $(X, \phi,s)$ space.}  Each line corresponds to a different
  system with $Z(\mu)$ as indicated.  The projections in the
  $(\phi,s)$ and $(X, s)$ planes show that the RCP $(X=0)$ is less
  disordered than the RLP $(X\to \infty)$.  The projection in the $(X,
  \phi)$ plane resembles qualitatively the compaction curves of the
  experiments \cite{chicago2,swinney,sdr}.  }
\label{entropy}
\end{figure}

\begin{figure}
  \centering { \vbox{ \resizebox{8cm}{!}{\includegraphics{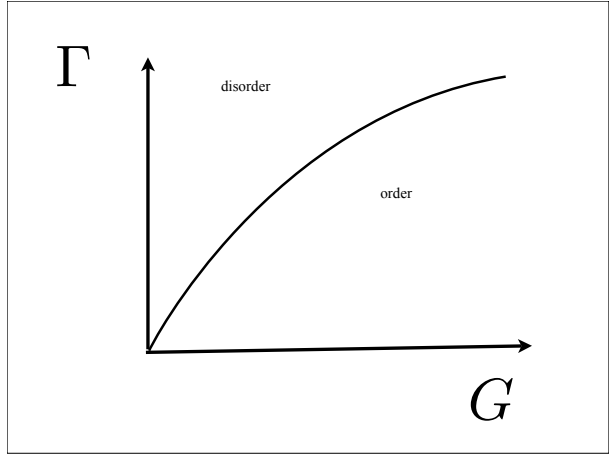}} }
  }
\caption{Dimensional analysis indicates that for finite shear modulus
  $G$ of the particles, the system can crystallize or randomize
  according to the quenched rate $\Gamma$. This argument allows for a
  comparison with hard-sphere simulations that are done in the limit
  $G\to \infty$.  In this case only $\Gamma\to \infty$ avoids partial
  crystallization.}
\label{G}
\end{figure}

\begin{figure}
  \centering {
    \resizebox{8cm}{!}{\includegraphics{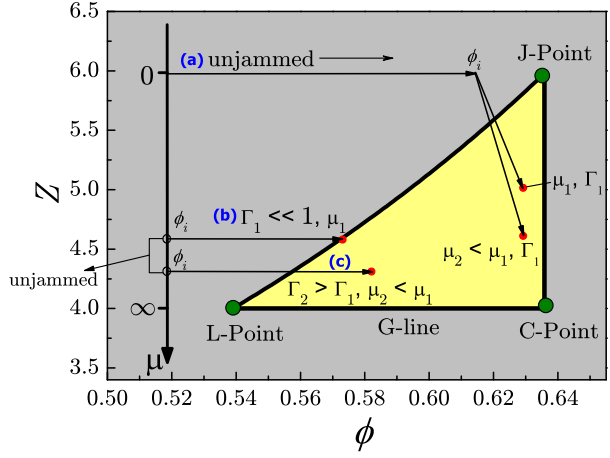}} }
  \caption{Schematic representation of the protocol used to
    dynamically achieved the packings in the phase diagram.  (a) The
    region near the C-point (high friction, high volume fraction) is
    the most difficult to access. Here, we prepare an unjammed sample
    with no friction at $\phi_i$ (=0.62 in the diagram) very close to
    RCP, and then we switch friction to compress to the required
    frictional jammed state. The packing follow the paths indicated in
    the figure for two friction coefficients.  On the other hand, the
    RLP-line is easily obtained by compressing unjammed packings with
    low $\phi_i$ (=0.52 in the diagram) at (b) a given compression
    rate $\Gamma_1$ for a given friction coefficient. (c) There is a
    small dependence on $\Gamma$, the faster the compression the
    deeper we enter in the phase diagram.}
  \label{protocol}
\end{figure}

\begin{figure}
\centering {
\vbox{
 \resizebox{8cm}{!}{\includegraphics{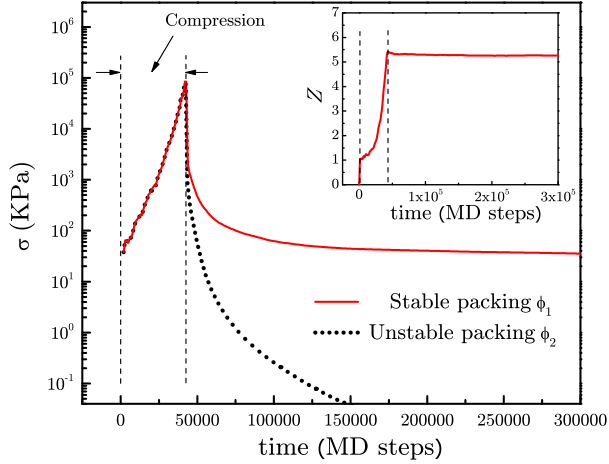}}
}
}
\caption{ Time evolution of stress (the pressure in the
    system) for two packings simulated as explained in the text.  The
    solid red line represents a packing with $\phi_{1}>\phi_{c}$ and
    dotted black line represents a packing with $\phi_{2}<\phi_{c}$,
    where $\phi_{1}=\phi _{2}+2\times10^{-4}$. The inset shows the
    time evolution of the coordination number.}
\label{criticality}
\end{figure}

\begin{figure}
\centering {
\resizebox{8cm}{!}{\includegraphics{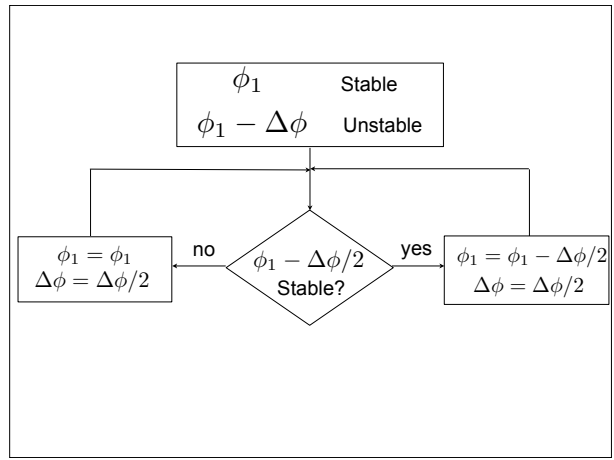}}
}
\caption{Flow chart for searching procedure of critical volume
  fraction $\phi_{c}$.} \label{chart}
\end{figure}

\begin{figure}
{ \vbox{ \resizebox{9cm}{!}{\includegraphics{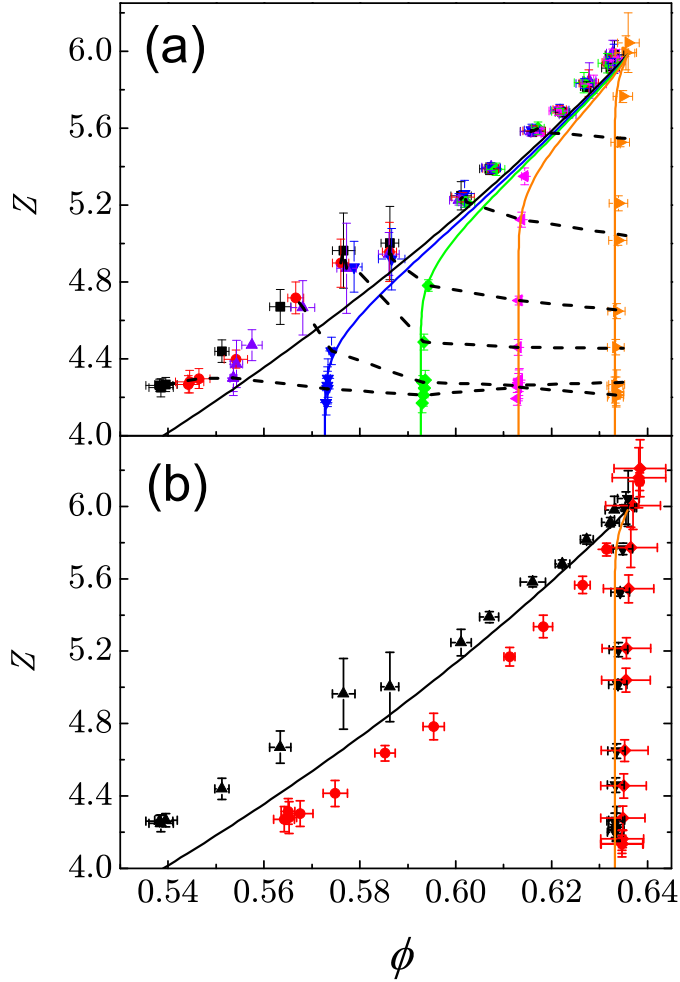}} } }
\caption{{\bf Phase diagram of jammed matter: Simulations}.  Numerical
  simulations demonstrate how to dynamically access the theoretically
  found states. The numerical protocol is parameterized by $(\phi_i,
  \Gamma, \eta,\mu)$.  (a) The plot shows the dependence of the final
  jammed states $(\phi,Z)$ on $\phi_i$ for a fix $\Gamma=10^{-7}$ and
  $\eta=10^{-3}$ (except for the orange $\blacktriangleright$ curve
  which is for $\eta=10^{-4}$) from left to right $\phi_i = 0.40$
  (black $\blacksquare$), 0.53 (red $\bullet$), 0.55 (violet
  $\blacktriangle$), 0.57 (blue $\blacktriangledown$), 0.59 (green
  $\blacklozenge$), 0.61 (pink $\blacktriangleleft$) and 0.63 (orange
  $\blacktriangleright$).  Each equal-color line set represents a
  different $\phi_i$ and the dashed lines join systems with the same
  friction $\mu$. Solid lines represent the theoretical results with
  $h_z=e^{-100}$ or $z_c=0.01$ for different compactivities measured
  in units of $10^{-3} V_g$. From left to right $X=\infty$ (black
  solid line), 1.62 (blue solid line), 1.38 (green solid line), 1.16
  (pink solid line) and 0.88 (orange solid line).  The error bars
  correspond to the s.d. over 10 realizations of the packings.  (b)
  The plot focuses on the dependence of $(\phi,Z)$ on $(\Gamma,\eta)$
  for two different $\phi_i$.  Solid black symbols are for
  $\phi_i=0.40$ and $(\Gamma,\eta)= (10^{-7},10^{-3})$ (black
  $\blacksquare$) and $(10^{-4},10^{-4})$ (black $\bullet$).  Open red
  symbols are for $\phi_i=0.63$ and $(\Gamma,\eta)= (10^{-7},10^{-3})$
  (red $\square$) and $(10^{-3},10^{-4})$ (red $\circ$).  }
\label{phaseb}
\end{figure}

\begin{figure}
  \centering {
    \resizebox{5cm}{!}{\includegraphics{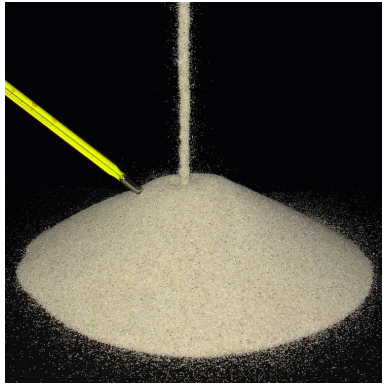}} }
  \caption{Measuring the temperature of sand. Pictorial
    representation.}
\label{thermometer}
\end{figure}

\begin{figure}
\centering {
\resizebox{8cm}{!}{\includegraphics{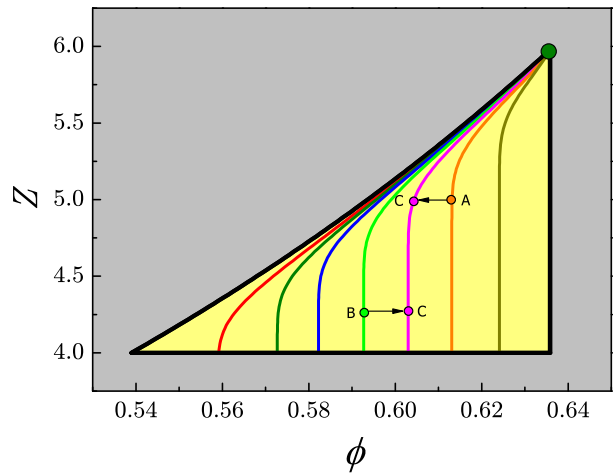}} }
\caption{ Sketch depicting a possible ABC experiment to test the
  zero-th law of granular thermodynamics and the validity of the phase
  diagram for finite compactivities.  For intermediate frictions, the
  A and B systems should follow the arrows to C when put into contact
  and allow to interchange volume by gently shaking.}
\label{abc}
\end{figure}

\begin{figure}
\centering \resizebox{8cm}{!}{\includegraphics{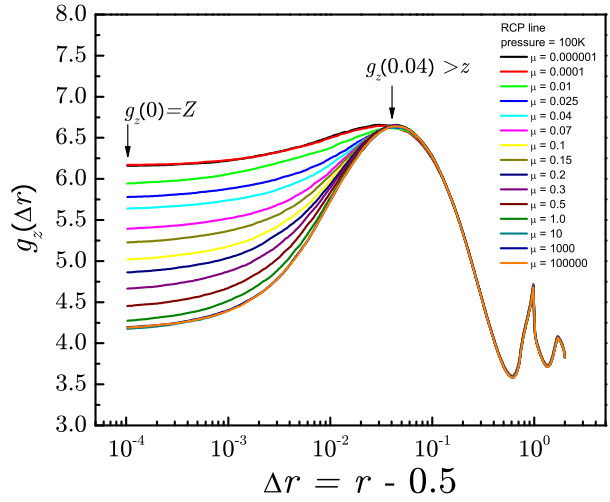}}
\caption{$g_z(\Delta r)$ of packings with various friction coefficient
  $\mu$ along RCP line. We set $2R=1$.} \label{gzr_RCP}
\end{figure}

\begin{figure}

\centering \resizebox{8cm}{!}{\includegraphics{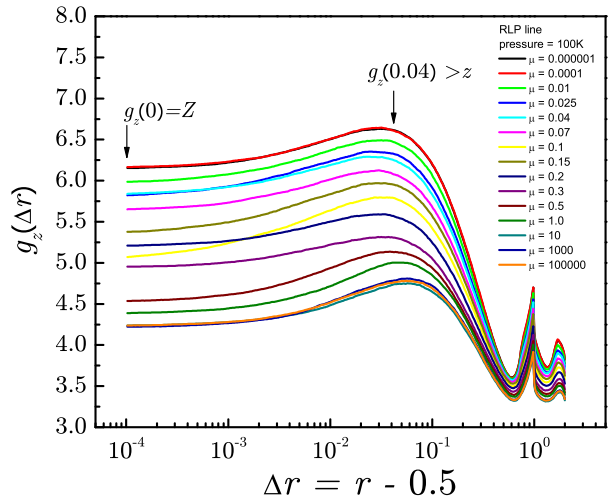}}
\caption{$g_z(\Delta r)$ of packings with various friction coefficient
$\mu$ along RLP line.} \label{gzr_RLP}
\end{figure}

\begin{figure}
\centering \resizebox{9cm}{!}{\includegraphics{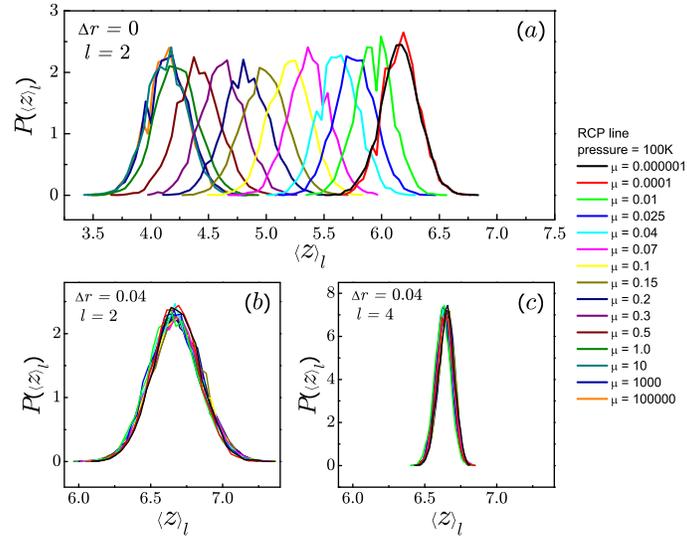}}
\caption{PDF of the coarse-grained coordination number $\langle
z\rangle_l$ for packings with various friction coefficient $\mu$ along
the RCP line. (a) $\Delta r=0$ and $l = 2$; (b) $\Delta r=0.04$ and $l =
2$; (c) $\Delta r=0.04$ and $l = 4$. } \label{cgz_RCP}
\end{figure}

\begin{figure}
\centering \resizebox{9cm}{!}{\includegraphics{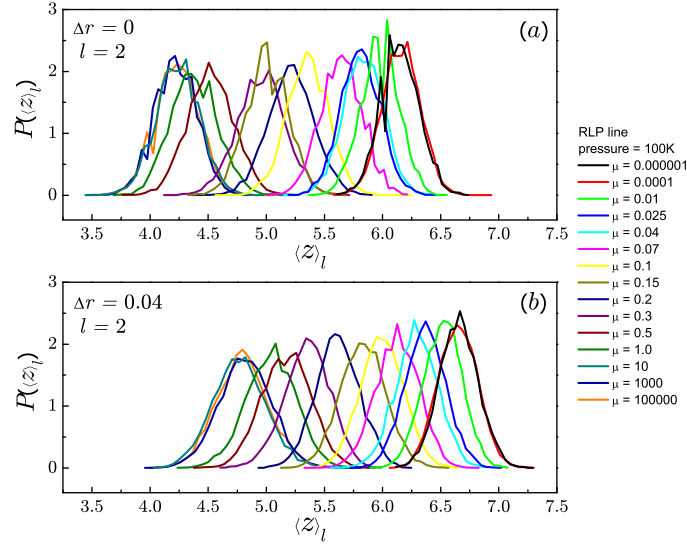}}
\caption{PDF of the coarse-grained coordination number $\langle
z\rangle_l$ for packings with various friction coefficient $\mu$ along
the RLP line. (a) $\Delta r=0$ and $l = 2$; (b) $\Delta r=0.04$ and $l =
2$.} \label{cgz_RLP}
\end{figure}

\begin{figure}
\centering \resizebox{9cm}{!}{\includegraphics{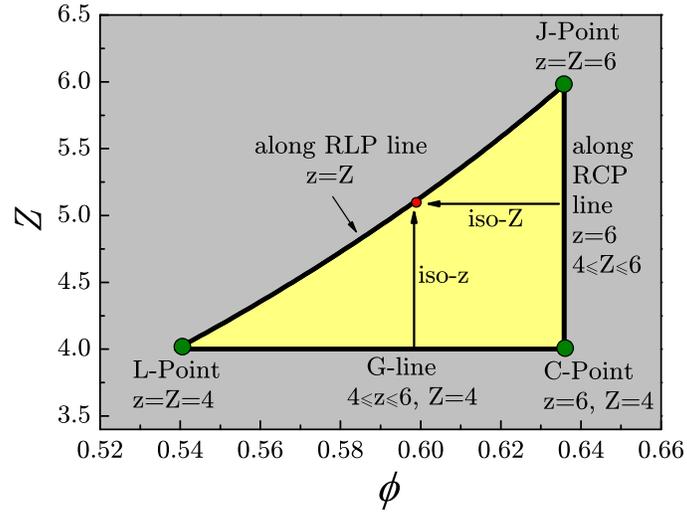}}
\caption{ Summary of the theoretical findings regarding the range of $z$ and $Z$ along the different iso-$z$, iso-$Z$, and iso-$X$ lines and the J, C, and L-points in the phase diagram.}
\label{z-summary}
\end{figure}

Beyond the elucidation of some questions in the sphere packing
problem, other problems can now be addressed systematically from the
point of view of what we have learned from the phase diagram. This
includes the investigation of the criticality of the jamming
transition from frictionless to frictional systems by extending the
phase space to $(\phi,Z,\sigma)$, and generating packings away from
the isostatic plane by isotropically compressing the packings
generated in the plane of Fig. \ref{phasea}.  Previous studies have
focused on the frictionless point and on fixed friction. According to
our results it is important to sample all the phase space for
different friction and compactivities according to the preparation
protocols explained above.

\begin{figure}
\centering \resizebox{9cm}{!}{\includegraphics{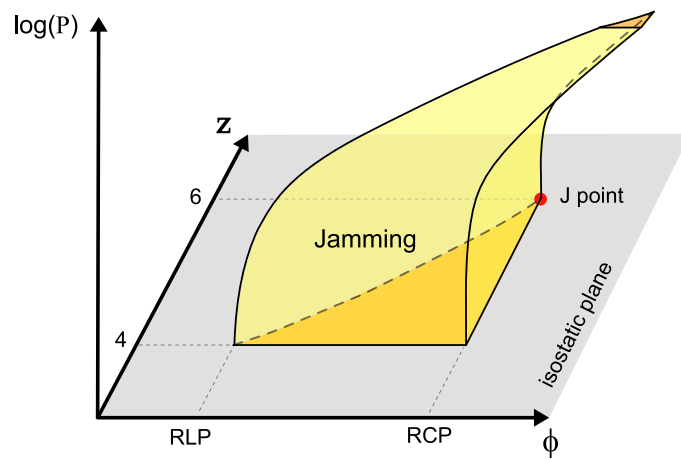}}
\caption{ Sketch depicting a possible extension of the phase diagram
    of Fig. \ref{phasea} in the $(\phi,Z)$ isostatic plane at the
    jamming transition to the full phase diagram in the
    $(\phi,Z,\sigma)$ space to investigate the criticality of the
    jamming transition for all packings with any friction
    coefficient.}
\label{phase2}
\end{figure}

Previous results find power law scaling
\cite{ohern,silbert,makse,zhang} where the stress vanishes as $\sigma
\sim (\phi-\phi_c)^{\alpha}$. A possible scenario is depicted in the
extended phase diagram of Fig. \ref{phase2} in the 3-dimensional space
$(\phi,Z,\sigma)$, which could be tested numerically and
experimentally.  We note in passing that, all the results in the
present study refer to the isostatic plane at the jamming transition
$\sigma=0$ or hard sphere limit. We expect that, as we compress
packings of soft spheres, different planes of $\sigma =$ constant $\ne
0$ will be obtained as shown in Fig. \ref{phase2}. These compressed
states are described not only by the compactivity $X$ but also the
angoricity $A$ through the full partition function in the VF ensemble,
Eq. (\ref{sw}).  Our preliminary results indicate that there seems to
be evidence of criticality as the isostatic plane is approached when
$\sigma \to 0$, although with exponents dependent on friction as well
as the value of the compactivity.

Additional topics to be addressed by means of the presented approach
include characterization of jamming in the phase space of
configurations, the problem of elasticity and Green's function and the
study of the pair correlation function $g(r)$, the volume distribution
$P({\cal W}_i^s)$, Voronoi volume, $P({\cal W}_i^{\rm vor})$, and
coordination number $P(z)$ , which are within the reach of the present
approach.  Other systems under consideration are two-dimensional
systems \cite{jamming5}, polydisperse systems \cite{Max},
as well as the mean field case of infinite dimension.  These are
interesting cases, since there are many questions debated as to what
is the more dense state in higher dimension.  Extensions to other
systems such as anisotropic particles are also within the reach of the
present approach.

Finally, a complete characterization of the VF-ensemble can now be
performed through the measurement of other quantities as a function of
the volume fraction, such as the force distribution $P(F_n,F_t)$, and
equations of state through the angoricity in the $(\phi,Z,\sigma)$
space.  Some of these quantities can be obtained analytically from the
ensembles allowing for interesting predictions.

The phase diagram introduced here serves as a beginning to understand
how random packings fill space in 3d.  The comparative advantage of
the present approach over extensive work done in the past, is in the
classification of all packings through $X$, $Z$ and $\phi$ in the
theoretical phase diagram from where these studies could be
systematically performed. This classification guides the search for
indications of jamming from a systematic point of view, through the
exploration of all jammed states from $\mu=0$ to $\mu\to \infty$.

{\bf Acknowledgements}. This work is supported by the National Science
Foundation, CMMT Division and the Department of Energy, Office of
Basic Energy Sciences, Geosciences Division. We are grateful to
B. Bruji\'c and A. Yupanqui for inspirations, C. Briscoe for a
critical reading of the manuscript and the hospitality of UFC,
Fortaleza where part of this work was done.

\clearpage


\clearpage

\end{document}